\documentclass[aps,amsmath,amssymb,amsfonts,preprint,nofootinbib]{revtex4}

\usepackage{graphicx}

\usepackage{tikz}
\usepackage{caption}
\usepackage{subcaption}
\usepackage{mathtools}

\usetikzlibrary{shapes.misc}
\usetikzlibrary{arrows.meta}
\tikzset{cross/.style={cross out, draw=black, minimum size=2*(#1-\pgflinewidth), inner sep=0pt, outer sep=0pt},
cross/.default={1pt}}
\tikzstyle{axisarrow} = [-{Latex[inset=0pt,length=5pt]}]

\newcommand{\RR}{{\mathbb R}}
\newcommand{\ra}{\rightarrow}
\newcommand{\eps}{\epsilon}

\newcommand{\Tr}{{\rm Tr}}

\newcommand{\br}{{\bf r}}

\newcommand{\ma}{{\mathcal M}}
\newcommand{\mb}{{\mathcal M'}}
\newcommand{\mc}{{\mathcal M''}}
\newcommand{\da}{{\delta\alpha}}
\newcommand{\dg}{{\delta\gamma}}

\newcommand{\R}{R}

\usepackage{amsmath}

\begin{document}

\title{Thermal Hall conductance and a relative topological invariant of gapped two-dimensional systems}
\author{Anton Kapustin}
\email{kapustin@theory.caltech.edu}
\author{Lev Spodyneiko}
\email{lionspo@caltech.edu}
\affiliation{California Institute of Technology, Pasadena, CA 91125, United States}

\begin{abstract}
We derive a Kubo-like formula for the thermal Hall conductance of a 2d lattice systems which is free from ambiguities associated with the definition of energy magnetization.
We use it to define a relative topological invariant of gapped 2d lattice systems at zero temperature. Up to a numerical factor, it can be identified with the difference of chiral central charges for the corresponding edge modes. This establishes the  bulk-boundary correspondence for the chiral central charge.
We also show that for any Local Commuting Projector Hamiltonian the relative chiral central charge vanishes, while for free fermionic systems it is related to the zero-temperature electric Hall conductance via the Wiedemann-Franz law.
\end{abstract}

\maketitle

\section{Introduction}\label{sec:intro}

There has been much theoretical as well as experimental interest in the thermal Hall effect. Just to give a couple of recent examples: (1) thermal Hall effect has been used to probe the non-Abelian nature of the $\nu=5/2$ FQHE state \cite{FQHE52}; (2) an unusual behavior of thermal Hall conductivity at low temperatures was observed in cuprate superconductors in the pseudogap region \cite{cuprates}.

Despite many theoretical works on the thermal Hall effect (see e.g. \cite{Cooperetal,Niuetal,Stone,BradlynRead,GeracieSon}), there are still unresolved issues with the very definition of thermal Hall conductivity. In fact, all known approaches to defining thermal Hall conductivity as a bulk property are plagued with ambiguities. To see what the issue is in the simplest possible setting, consider a macroscopic system where the only conserved quantity carried by the low-energy excitations is energy (for example, an insulator at temperatures well below the band gap). One could expect that thermal Hall conductivity appears as a transport coefficient in the hydrodynamic description, but this is not the case: there is no physical time-reversal-odd transport coefficient at leading order in the derivative expansion. The conservation law for the energy density $\eps$ is
\begin{equation}
\frac{\partial \eps}{\partial t}=-\nabla\cdot 
{\bf j}^E.
\end{equation}
In the hydrodynamic limit one can expand the energy current ${\bf j}^E$ to first order in derivatives of $\eps$, or equivalently in derivatives of the temperature $T$:
\begin{equation}
{\bf j}^E_m=-\kappa_{m\ell}(T) \partial_\ell T .
\end{equation}
Hence the conservation law becomes
\begin{equation}
c(T)\frac{\partial T}{\partial t}=\kappa_{m\ell}\partial_m\partial_\ell T+\kappa'_{m\ell}\partial_m T\, \partial_\ell T,
\end{equation}
where $c(T)$ is the heat capacity and the prime denotes derivative with respect
to $T$. The r.h.s. of this equation depends only on the symmetric part $\kappa^S_{m\ell}$ of the tensor $\kappa_{ml},$ which by Onsager reciprocity is the same as its time-reversal-even part. The anti-symmetric part $\kappa_{ml}^A$ has no observable effect in the bulk. While the energy current through a surface (or, in the 2d context, through a line) depends on the whole tensor $\kappa_{m\ell}$, the contribution of $\kappa_{m\ell}^A$ can be thought of as a boundary effect. Indeed, if we define 
\begin{equation}
\beta^A_{m\ell}(T)=\int^T \kappa^A_{m\ell}(u) du,
\end{equation}
then in 3d the Stokes' theorem gives
\begin{equation}
-\int_\Sigma d{\bf \Sigma}_m \kappa^A_{m\ell} \partial_\ell T \  =-\frac12\int_{\partial\Sigma} d{\bf l}_k \eps_{kpm} \beta^A_{pm}(T).
\end{equation}
Similarly, in 2d the contribution of $\kappa^A_{m\ell}$ to the energy current through a line can be written as a boundary term.
The conclusion seems to be that thermal Hall conductivity has no meaning as a bulk transport property, either in 3d or 2d. One manifestation of this is that Kubo-type formulas for thermal Hall conductivity are ambiguous: they involve ``energy magnetization'', which is defined only up to an arbitrary function of temperature and other parameters \cite{Cooperetal,Niuetal,GeracieSon}. This leaves us with the question of how to describe theoretically the thermal Hall conductivity measured in experiments.

In the 2d case the tensor $\kappa^A_{m\ell}$ reduces to a single quantity \footnote{We use the notation $\kappa^A$ instead of the more standard $\kappa_{xy}$ to avoid confusion with the off-diagonal component of $\kappa^S$ which may be nonzero if rotational invariance is broken.}, the thermal Hall conductivity  $\kappa^A=\frac{1}{2}\eps_{m\ell}\kappa^A_{m\ell}$, and there is an alternative line of reasoning which suggests that in certain circumstances $\kappa^A$  {\it can} be defined in bulk terms. Consider a material with a bulk energy gap. There might still be gapless excitations at the edges, and we will assume that they are described by a 1+1d Conformal Field Theory. Then it seems natural to relate $\kappa^A(T)$ to the chiral central charge of the edge CFT:
\begin{equation}\label{cardy}
\kappa^A(T)\simeq\frac{\pi T}{6}(c_R-c_L).
\end{equation}
To see why this is natural, recall that a chiral 1+1d CFT at temperature $T$ carries an equilibrium energy current $I^E=\frac{\pi T^2}{12}(c_R-c_L)$ \cite{CFT1,CFT2}. Thus in a strip of a 2d material whose boundaries are kept at temperatures $T$ and $T+\Delta T$, where $T$ is much smaller than the bulk energy gap and $\Delta T\ll T$, there is a net energy current
\begin{equation}
I^E\simeq \frac{\pi T} {6}(c_R-c_L)\Delta T.
\end{equation}
If we define $\kappa^A=I^E/\Delta T$, we get  (\ref{cardy}). 

On the other hand, it has been shown in \cite{energyBloch} that the chiral central charge of the edge modes (and more generally, the equilibrium energy current carried by the edge modes) is independent of the particular edge. Hence the low-temperature  thermal Hall conductivity of a gapped 2d material defined via (\ref{cardy}) is a well-defined bulk property.\footnote{For gapped 2d systems at low temperatures, one can also try to define thermal Hall conductivity as the coefficient of the gravitational Chern-Simons term in the low-energy effective action \cite{gravCSone,gravCStwo,gravCSthree}. As explained in \cite{Stone}, the energy current corresponding to the gravitational Chern-Simons term is of higher order in derivatives, in agreement with the above discussion. However, there is no natural way to couple a typical condensed matter system to gravity, therefore this prescription is ambiguous. } 

The results of \cite{energyBloch} also imply that the chiral central charge of the edge modes does not vary as one changes the parameters of the Hamiltonian. Therefore the low-$T$ thermal Hall conductivity is a topological invariant of the gapped 2d material. Finally, the above arguments make no assumption about the way the temperature varies within the strip. Thus the low-temperature thermal Hall {\it conductance} of a strip of a gapped 2d material coincides with its thermal Hall conductivity and is a well-defined bulk property as well. One does not expect this to hold at arbitrary temperatures, or for gapless materials at low $T$. 

This leads us to ask the following questions.

Q1. Is the thermal Hall conductivity measured in experiments (at general temperatures) a well-defined bulk quantity? If yes, then how is this compatible with the above arguments that thermal Hall conductivity is not a well-defined bulk transport coefficient?

Q2. Is there a microscopic Kubo-type formula for the thermal Hall conductance and conductivity measured in experiments (at general temperatures) which makes no reference to the choice of the edge?

Q3. Is it true that the low-temperature thermal Hall conductance of a gapped 2d material is independent of the detailed shape of the temperature profile and thus coincides with the thermal Hall conductivity even if the edge is not described by Conformal Field Theory?

Q4. Is it true that the low-temperature thermal Hall conductance of a gapped 2d material is linear in $T$ at low $T$ even if the edge is not described by Conformal Field Theory?

Q5. Is it true that the low-temperature thermal Hall conductance of a gapped 2d material is a topological invariant of the phase, in the sense that it does not change when the parameters of the Hamiltonian vary without crossing a bulk zero-temperature phase transition?

The goal of this paper is to provide answers to the above questions in the case of lattice 2d systems. We show (with varying degree of rigor) that the answers to all these questions is "yes". In addition, we show that for systems described by Commuting Projector Hamiltonians thermal Hall conductance vanishes identically for all temperatures. We also show that for 2d gapped free fermionic systems of class A (that is, for non-interacting possibly disordered 2d insulators) thermal Hall conductance at low temperatures and electric Hall conductance are related via the Wiedemann-Franz law.

\section{Summary of results}

Our main observation is that while it is problematic to give a definition of thermal Hall conductance which is not "contaminated" with edge effects, there is no such difficulty for derivatives of the thermal Hall conductance with respect to parameters of the Hamiltonian. We derive microscopic Kubo-type formulas for all such derivatives in a straightforward manner. A limitation of such formulas is that they hold only away from phase transitions. This is a common limitation of the usual linear response theory which assumes that correlations are short-range in order to be able to make a derivative expansion.  

Kubo-like formulas for the derivatives of the thermal Hall conductance can be used to compute the difference of thermal Hall conductances $\kappa^A_{\ma\mb}=\kappa^A_{\ma}-\kappa^A_{\mb}$ of two 2d  materials $\ma$ and $\mb$. One chooses a path in the parameter space connecting the two Hamiltonians and avoiding bulk phase transitions and integrates the derivative along this path. Specializing to a linear temperature profile, we also get a formula for the difference of thermal Hall {\it conductivities}. 

Our Kubo-like formula satisfies an important consistency check: the integral defining the "relative thermal Hall conductance" $\kappa^A_{\ma\mb}$ is independent of the choice of the path. We give both an intuitive argument based on the absence of macroscopic energy currents in equilibrium (which has been proved recently \cite{energyBloch}) and a more formal mathematical argument for lattice systems. This allows us to standardize the choice of paths used to compute $\kappa^A_{\ma\mb}$. For example, for lattice systems with a finite-dimensional on-site space of states (such as fermion systems and spin systems) one can use paths which pass through the infinite-temperature phase. Since the infinite-temperature phase is the same for all lattice Hamiltonians, this makes it more plausible that a suitable path can be found for all pairs of materials $\ma,\mb$. 

One can interpret the integral formula for the relative thermal Hall conductance in more physical terms if one considers a smooth edge between the materials $\ma$ and $\mb$ which interpolates between the two Hamiltonians in the physical space. If one applies linear response theory to this system and assumes that the temperature gradient is negligible in the edge region, one gets precisely our integral formula. Path-independence of the integral formula is then equivalent to the independence of the choice of the edge between $\ma$ and $\mb$. The latter property can be traced again to the absence of macroscopic energy currents in equilibrium. 

This physical interpretation clarifies why it is not possible to write a well-defined microscopic formula for $\kappa^A_\ma$ even though it is possible to write down such a formula for the electric Hall conductance $\sigma^A_{\ma}$ of a single material $\ma$. In the electric case, torus geometry provides a theoretical set-up where $\sigma^A_{\ma}$ can be measured without introducing edges. In this geometry, electric field is created using a time-dependent vector potential rather than a scalar potential. There is no thermal analogue of the torus set-up, and this is why only the relative thermal Hall conductance $\kappa^A_{\ma\mb}$ of two materials has a physical significance.

In most experiments, one of the materials is the vacuum and the difference between  the thermal Hall conductances  of the material and the vacuum is measured. If one normalizes the thermal Hall conductance of the vacuum to be zero, then the thermal Hall conductance of a material $\ma$ relative to the vacuum can be declared to be the ``absolute'' thermal Hall conductance of $\ma$. Nevertheless, it is important to keep in mind that this is just a normalization condition, not something forced on us by physics.\footnote{This was first noticed by H. Casimir in his landmark paper on Onsager reciprocity \cite{Casimir}. Casimir showed that invariance under time-reversal, strictly speaking, does not require the anti-symmetric part of the thermal conductivity tensor to vanish. Vanishing is only obtained if one normalizes the thermal Hall conductivity of the vacuum to be zero.} One consequence of this is that there is no microscopic formula for the thermal Hall conductance which is local in the parameter space (that is, depends only on correlators for a particular Hamiltonian).

The results described above answer questions Q1 and Q2. Specifically, although thermal Hall conductivity is not a well-defined  bulk transport coefficient and can be measured only in the presence of an edge or another inhomogeneity, thermal Hall energy flux can be shown to be independent of the properties of the edge, provided the variation of the temperature on the length scale determined by the edge is negligible. 

To answer Q3, Q4 and Q5 we study the low-temperature behavior of our formula for $\kappa^A_{\ma\mb}$. Using the same method as in the work of Niu and Thouless on the electric Hall conductance \cite{NiuThouless}, we show that the low-$T$ behavior of $\kappa^A_{\ma\mb}$ is independent of the precise temperature profile, up to terms exponentially suppressed in the temperature. This answers Q3. We also argue that that derivatives of the thermal Hall conductance of a gapped 2d system with respect to parameters of the Hamiltonian are exponentially small for low $T$ if there is a bulk energy gap. This answers Q5. Then we explain how to include the temperature $T$ among the parameters and argue that the $T$-derivative of the dimensionless quantity $\kappa^A_{\ma\mb}/T$ is also exponentially small at low $T$ if there is an energy gap. This implies that $\kappa^A_{\ma\mb}$ is linear in $T$ up to exponentially small corrections. This answers Q4. Together with Q5, this shows that the coefficient of the $T$-linear term in $\kappa^A_{\ma\mb}$ is a topological invariant of the phase. 

In this paper we focus on lattice 2d systems. This allows to give a completely general formula for derivatives of the thermal Hall conductance with respect to arbitrary parameters of the Hamiltonian. However, working with lattice systems leads to certain technical complications. In particular, when working with currents on a lattice it is very convenient to make use of some mathematical machinery which is not familiar to most physicists, such as the Vietoris-Rips complex. Without this machinery, computations become very obscure. To make the paper more accessible, we relegate most mathematical details to appendices. 

Since the definition of thermal Hall conductance is rather subtle, we begin with a discussion of the electric Hall conductance. Some of the subtleties arise already in this context. Then we move on to the thermal case and derive a Kubo-like formula for derivatives of the thermal Hall conductance with respect to parameters. We argue that the integral defining the difference of thermal Hall conductances of two materials is independent of the path used to compute it. Then we discuss the low-temperature behavior of the thermal Hall conductance and show that for gapped systems it is linear in $T$ up to exponentially small corrections and that its slope  is a topological invariant of the phase.  We also show that that for systems described by Local Commuting Projector Hamiltonians thermal Hall conductance vanishes identically. Therefore such systems cannot have edge modes described by a CFT with a nonzero chiral central charge. This is an energy counterpart of the recently proved result that in such systems the zero-temperature electric Hall conductance vanishes \cite{KapFid}. In one of the appendices, we show by a direct computation that for gapped free fermionic systems of class A the relative thermal Hall conductance of the $T=0$ and $T=\infty$ states is related to the zero-temperature electric Hall conductance through a version of the Wiedemann-Franz law. The derivation does not assume translational invariance. Other appendices set up the mathematical machinery mentioned above and supply some details of the derivation.

We thank Yu-An Chen for participation in the early stages of this work and M. Hastings, H. Watanabe, A. Kitaev, and H. Edelsbrunner for discussions.
This research was supported in part by the U.S.\ Department of Energy, Office of Science, Office of High Energy Physics, under Award Number DE-SC0011632. A.K. was also supported by the Simons Investigator Award.

\section{Electric Hall conductance}

\subsection{Electric currents on a lattice}

A lattice system  in $d$-dimensions has a Hilbert space $V=\otimes_{p\in\Lambda} V_p$, where $\Lambda$ (``the lattice'') is a uniformly discrete subset of $\RR^d$ (that is, there is a minimal distance $D>0$ between all points), and all $V_p$ are  finite-dimensional. An observable is  localized at a point $p\in\Lambda$ if it has the form $A\otimes 1_{\Lambda\backslash p}$ for some $A: V_p\ra V_p$. An observable is localized on a subset $\Lambda'\subset\Lambda$ if it commutes with all observables localized at any  $p\notin \Lambda'$. A local observable $A$ is an observable localized on a finite set $\Lambda'\subset\Lambda$, which will be called the support of $A$.

Hamiltonian of a lattice system has the form 
\begin{equation}\label{latticeH}
H=\sum_{p\in\Lambda} H_p,
\end{equation}
where the operators $H_p:V\ra V$ are Hermitian and local. We will assume that the Hamiltonian has a finite range $\R$, which means that each $H_p$ is a local observable supported in a ball of radius $\R$ centered at $p$. This implies that $[H_p,H_q]=0$ whenever $|p-q|>2\R$. We will also assume that the operators $H_p$ are uniformly bounded, i.e. there exists $C>0$ such that $||H_p||<C$ for all $p\in\Lambda$.

To define electric currents, we assume that the system has an on-site $U(1)$ symmetry. Thus we are given a $U(1)$ action on each $V_p$, with the generator $Q_p$ (a Hermitian operator on $V_p$ with integral eigenvalues). The total $U(1)$ charge is $Q_{tot}=\sum_{p\in\Lambda} Q_p$. Further, we assume that  $[Q_{tot},H_p]=0$ for any $p\in\Lambda$. Since the time derivative of $Q_q$ is
\begin{equation}\label{electric conservation law}
\frac{dQ_q}{dt}=i\sum_{p\in\Lambda} [H_p,Q_q],
\end{equation}
it appears natural to define the $U(1)$ current from $q$ to $p$ by $J_{pq}=-i[H_p,Q_q]$. However, this does not satisfy a physically desirable property $J_{qp}=-J_{pq}$. Instead we define
\begin{equation}\label{electric current}
J_{pq}=i[H_q,Q_p]-i[H_p,Q_q].
\end{equation}
The lattice current thus defined satisfies $J_{qp}=-J_{pq}$ as well as
\begin{equation}\label{electricconserv}
\frac{d Q_q}{dt}=-\sum_{p\in\Lambda} J_{pq}.
\end{equation}
Each of the operators $J_{pq}$ is local in the above sense (it commutes with operators whose supports are sufficiently far from both $p$ and $q$). But the collection of all $J_{pq}$ is also local in a different sense: $J_{pq}$ vanishes when $|p-q|$ is sufficiently large (specifically, greater than $\R$). Objects depending on two or more points of $\Lambda$ which vanish when the any of the two points are sufficiently far will be called finite-range. So one can also say that the current $J_{pq}$ is finite-range. The property of being finite-range makes sense not just for operators, but also for c-number quantities depending on several points of $\Lambda$. 

While the above definition of the electric current seems natural, it is not completely unique. Let $U_{pqr}$ be any function of three points which takes values in local operators, is skew-symmetric in all three variables, and is finite-range. If we define 
\begin{equation}\label{electric_current_ambiguity}
J'_{pq}=J_{pq}+\sum_r U_{pqr},
\end{equation}
then it is easy to see that $J'_{pq}$ satisfies the same requirements as $J_{pq}$ and therefore is also a physically acceptable current. This is a lattice counterpart of the continuum statement that only $\nabla\cdot {\bf j}$ has a physical significance, and thus one can replace ${\bf j}\mapsto {\bf j}+\nabla\times {\bf u}$, where ${\bf u}$ is arbitrary, without affecting any physical predictions. In the lattice case, it is not obvious that the only ambiguity in the definition of the current is (\ref{electric_current_ambiguity}). This is shown in Appendix \ref{appendix:math} under some natural assumptions on $\Lambda$. 

Suppose $\Lambda$ is decomposed into a disjoint union of two sets, $\Lambda=A\cup B$, $A\cap B=\emptyset .$ The current from $B$ to $A$ is defined as
\begin{equation}
J(A,B)=\sum_{p\in A}\sum_{q\in B} J_{pq}.
\end{equation}
It is not difficult to check that $J(A,B)$ does not change if one replaces $J_{pq}$ with $J'_{pq}$ defined in (\ref{electric_current_ambiguity}). This is because $J(A,B)$ is physical: it is equal to minus the rate of change of the electric charge in region $B$. This is expressed by the equation
\begin{equation}
\frac{dQ(B)}{dt}=-J(A,B)
\end{equation}
Here $Q(B)=\sum_{p\in B} Q_p$. 

More generally, given a skew-symmetric function $\eta(p,q):\Lambda\times\Lambda\ra\RR$, one can define
\begin{equation}\label{jf}
J(\eta)=\frac12\sum_{p,q} \eta(p,q) J_{pq}.
\end{equation}
In general, this expression is not physical: it changes under the redefinition (\ref{electric_current_ambiguity}). However, if $\eta(p,q)$ satisfies
\begin{equation}
\eta(p,q)+\eta(q,r)+\eta(r,p)=0,\quad \forall p,q,r\in\Lambda,
\end{equation}
then one can check that $J(\eta)$ is invariant under substitutions (\ref{electric_current_ambiguity}) and thus is physical. Such checks become much easier if one uses the mathematical machinery explained in Appendix \ref{appendix:math}.
In the case $\eta(p,q)=\chi_B(q)-\chi_B(p)$, where $\chi_B(p)=1$ for $p\in B$ and $\chi_B(p)=0$ otherwise, $J(\eta)$ reduces to $J(A,B)$.

\subsection{Kubo formula for the electric Hall conductance}\label{sec:electricHall}

Usually, Kubo formula is written down for conductivity rather than conductance. That is, it is assumed that the electric field is uniform across all relevant scales. For our purposes, it will be important to have a formula for the electric Hall current which does not assume that the electric field is uniform. 

Consider a time-dependent perturbation of the Hamiltonian of the form
\begin{equation}
\Delta H(t)=\eps e^{st} \sum_{p\in\Lambda} g(p) Q_p,
\end{equation}
where the real parameter $\eps$ is small and $g: \Lambda\ra\RR$ is arbitrary for now. This perturbation corresponds to adiabatically switching on an electric potential $\eps g$. Assuming that at $t=-\infty$ the system  is in an equilibrium state at temperature $T$, at $t=0$ the system will be in a non-equilibrium steady state. The change in the expectation value of an observable $A$ at $t=0$ relative to the expectation value at $t=-\infty$ is given by the general Kubo formula
\begin{equation}\label{generalKubo}
\Delta \langle A\rangle=\eps \lim_{s\ra 0+} \beta\int_0^\infty e^{-st} \left\langle\left\langle A(t); \sum_p\frac{1}{i}[H,g(p) Q_p]\right\rangle\right\rangle  dt.
\end{equation}
Here Heisenberg-picture operators are  defined as usual, $A(t) = e^{i H t} A e^{-i H t},$ and double brackets $\langle \langle \dots\rangle \rangle$ denote Kubo's canonical pairing, see Appendix B. We also assumed that $A$ doesn't have an explicit dependence on $\eps$. 

For an infinite system, the existence of the limit $s\ra 0+$ in eq. (\ref{generalKubo}) is far from obvious. When both the perturbation $\Delta H$ and the observable $A$ are supported on a compact set $K\subset\Lambda$, the existence of the limit has been proved in \cite{linear response and KMS}. When $g$ is nonzero only on a compact set $K\subset\Lambda$, but $A$ is supported on a non-compact set, we still expect the limit to exist, at least away from phase transitions. Indeed, if the correlation length is finite, the state of the system far from $K$ is unaffected by the perturbation, and we can effectively truncate the support $A$ to be compact, thereby reducing to the case when both $g$ and $A$ are compactly supported. More generally, when the intersection of the supports of $g$ and $A$ is compact, the same argument suggests that $\Delta\langle A\rangle$ is well-defined.

\begin{figure}[ht]
    \centering
% Part (a)
  \begin{subfigure}[b]{0.3\textwidth}
    \caption{\qquad\qquad\qquad\qquad\qquad}
    \label{g kink fig}
    \centering
    \includegraphics{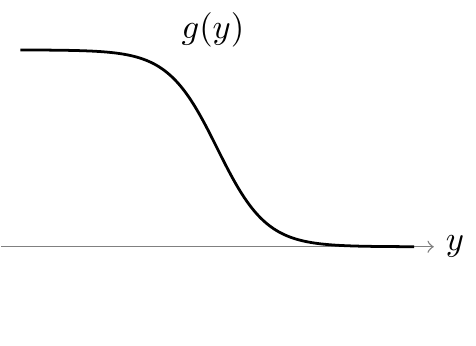}
\end{subfigure}
\hfill
% Part (b)
\begin{subfigure}[b]{0.3\textwidth}
  \centering
    \caption{\qquad\qquad\qquad\qquad}
    \label{g hat fig}
  \includegraphics{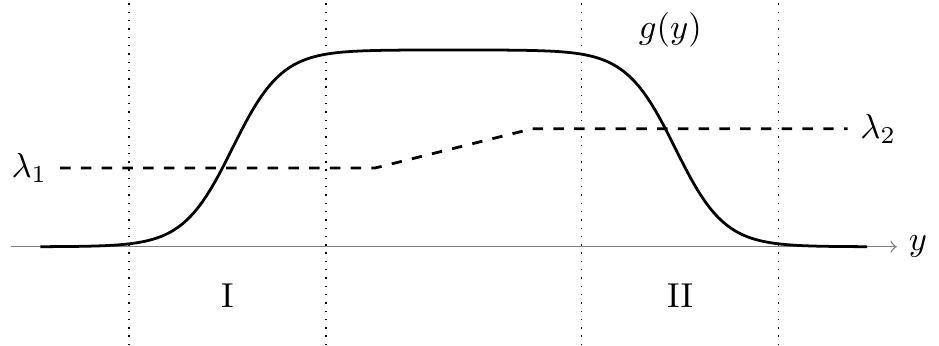}
\end{subfigure}
\hspace*{5cm} 
    \caption{}
\end{figure}

From now on we specialize to the 2d case, unless explicitly stated otherwise. To compute the quantum Hall conductance of an infinite 2d system, we would like $A$ to be the electric current across a vertical line $x=a$, and $g$ to be a function which depends only on $y$, vanishes at $y=+\infty$ and approaches $1$ at $y=-\infty$, see Fig. \ref{g kink fig}. Such a function $g$ corresponds to the net electric potential change $-\eps$ from $y=-\infty$ to $y=+\infty$. However, such $A$ and $g$ do not satisfy the condition on supports explained above. Another way to explain a potential problem is to note that while the electric field corresponding to such a function $g$ is vanishingly small for $y\ll 0$ and all $t$, the state of the system at $t=0$ and $y\ll 0$ is different from that at $t=-\infty$ and $y\ll 0$ because the electrochemical potential changes by $-\eps$. Since the expectation value of the current density is nonzero even in equilibrium and may depend on the electrochemical potential, the change in the current density between $t=0$ and $t=-\infty$ need not vanish at large negative $y$, and then the change in the net current across the line $x=a$ will be ill-defined.  

One way to avoid this difficulty is to make the $y$ direction periodic and to perturb the system by a constant vector potential rather than a scalar potential. However, this approach does not have an analog in the case of thermal transport, which is our primary interest. Alternatively, one can take $g$ to vanish both for $y\ll 0$ and $y\gg 0$. For example, one can take $g$ to look as in Fig. \ref{g hat fig}. Then the electric field is smooth in the regions $I$ and $II$ and has opposite magnitudes there. Elsewhere it is zero. If the system is homogeneous, the net electric Hall current in the $x$ direction will be zero. However, if the system is inhomogeneous, then the electric Hall conductance of the two regions may be different, and the net electric Hall current will be given by
\begin{equation}\label{definition of Hall conductance}
\Delta\langle A\rangle=-\eps\int_{-\infty}^{+\infty} \sigma_{xy}\, \partial_y g\, dy=\eps \int_{-\infty}^{+\infty} g\,  \partial_y\sigma_{xy} dx=\eps \int_{\lambda_1}^{\lambda_2}  \frac{\partial\sigma_{xy}}{\partial\lambda}\, d\lambda.
\end{equation}
Here we assumed that the system is homogeneous in regions $I$ and $II$, while in the intermediate region some parameter of the Hamiltonian $\lambda$ varies from $\lambda_1$ to $\lambda_2$ as $y$ is increased.
This approach allows one to compute the derivatives of the Hall conductance with respect to parameters. Integrating these derivatives along a path in the space of parameters, one can compute the relative electric Hall conductance of two systems, provided the path avoids phase transitions. This is good enough, since in practice one usually measures the relative electric Hall conductance of a particular material and vacuum.  

As discussed in the previous section, the net current through a vertical line $x=a$ is defined as 
\begin{equation}
J_a=\frac12\sum_{p,q} J_{pq}(f(q)-f(p)),
\end{equation}
where $f(p)=\theta(a-x(p))$ is a step-function. More generally, one can consider the expression (\ref{jf}) where one sets  $\eta(p,q)=f(q)-f(p)$ for some function $f:\Lambda\ra \RR$ which is equal to $1$ if $x(p)\ll 0$ and equal to $0$ if $x(p)\gg 0$. That is, $f$ is a smeared step-function in the $x$-direction. 

In what follows, we will use the following notation. Given any function $f:\Lambda\ra \RR$, we define a function $\delta f:\Lambda\times\Lambda\ra\RR$ by $(\delta f)(p,q)=f(q)-f(p)$. One can view  the operation $\delta$ as a lattice analogue of the gradient operator $\nabla$. For more details on this notaton see Appendix \ref{appendix:math}. Thus the smeared current (\ref{jf}) with $\eta(p,q)=f(q)-f(p)$ will be denoted $J(\delta f)$. While $J_a$ is the rate of change of the charge in the region $x>a$, $J(\delta f)$ is the minus the rate of change of the operator 
\begin{equation}
Q(f)=\sum_p f(p) Q_p
\end{equation}
That is,
\begin{equation}\label{rate of change}
i[H,Q(f)]=-J(\delta f).
\end{equation}

It is very important for what follows that when $f$ is a smeared step-function, $J(\delta f)$ is a local operator supported in a vertical strip on $\RR^2$, roughly where $f$ is neither $0$ nor $1$. Indeed, on the one hand, $J_{pq}$ is nonzero only if $|p-q|<R$. On the other hand, $f(q)-f(p)$ is zero if both $x(p)$ and $x(q)$ are sufficiently large and positive, as well as when both $x(p)$ and $x(q)$ are sufficiently large and negative. The combined effect of this is that $J(\delta f)$ is a sum of local operators supported in a vertical strip which is infinite in the $y$-direction but has a finite width in the $x$-direction.

Applying the general Kubo formula (\ref{generalKubo}) to $A=J(\delta f)$, we get 
\begin{equation}
\Delta\langle J(\delta f)\rangle=\eps \beta\lim_{s\ra 0+} \int_0^\infty  e^{-st} \langle\langle J(\delta f,t);J(\delta g)\rangle\rangle\, dt.
\end{equation}
Here we identified $\sum_p\frac{1}{i}[H, g(p) Q_p]$ with $J(\delta g)$ and denoted by $J(\delta f,t)$ the time-translation of $J(\delta f)$ by $t$. Note that $J(\delta g)$ is supported in a horizontal strip on $\RR^2$. More precisely, if $g$ is as in Fig. 1a, then $J(\delta g)$ is supported in a horizontal strip. If $g$ depends on $y$ as in Fig. 1b, then $J(\delta g)$ is supported in two horizontal strips corresponding to regions I and II in Fig. 1b. 

Recall now that we consider a Hamiltonian depending on a parameter $\lambda$ which varies with $y$ such that in region $I$ $\lambda=\lambda_1$, in  region $II$  $\lambda=\lambda_2$, and in the intermediate region $\lambda$ interpolates between these two values without crossing a phase transition. We assume that $\lambda_2-\lambda_1$ is small. We also choose $g$ as in Fig. \ref{g hat fig}. Then $J(\delta g)$ is a sum of operators supported in regions $I$ and $II$. We can make this explicit by writing $\delta g= \delta g_I+\delta g_{II}$, where $g_{II}$ interpolates between $0$ and $1$ as one moves from $y=+\infty$ to the intermediate region,  and $g_I$ interpolates from $1$ to $0$ as one moves from the intermediate region to $y=-\infty$. If the electric field in region $I$ is minus the translate of the electric field in region $II$, then $J(\delta g_I)$ is minus the translate of $J(\delta g_{II})$, and to linear order in $\Delta\lambda$ we get
\begin{equation}\label{DeltaJHall}
\Delta\langle J(\delta f)\rangle= \eps(\lambda_2-\lambda_1)\frac{\partial}{\partial\lambda}\left[ \beta\lim_{s\ra 0+} \int_0^\infty e^{-st} \langle\langle J(\delta f,t);J(\delta g_{II})\rangle\rangle dt\right]+O\Big((\lambda_2-\lambda_1)^2\Big).
\end{equation}
Here we implicitly assumed that the correlator 
\begin{equation}
\lim_{s\ra 0+}\int_0^\infty dt e^{-st}\langle\langle J(\delta f,t); J(\delta g)\rangle\rangle,
\end{equation}
depends only on the Hamiltonian in some neighborhood of the intersection of supports of $J(\delta f)$ and $J(\delta g)$, and thus when evaluating it one may assume that either $\lambda=\lambda_1$ or $\lambda=\lambda_2$. 

Comparing eq. (\ref{DeltaJHall}) with eq. (\ref{definition of Hall conductance}),  we get 
\begin{equation}
\frac{\partial \sigma_{xy}(f,g)}{\partial \lambda}=\frac{\partial}{\partial\lambda}\left[\beta \lim_{s\ra 0+}\int_0^\infty e^{-st} \langle\langle J(\delta f,t); J(\delta g)\rangle\rangle dt\right],
\end{equation}
where $g$ is now a function depending only on $y$ which interpolates between $1$ and $0$ as $y$ varies from $-\infty$ to $+\infty$, and $f(p)=\theta(a-x(p))$. 
This formula determines the electric Hall conductance up to an arbitrary constant. If we define the electric Hall conductance of vacuum to be zero, then we get a Kubo formula for the electric Hall conductance itself:
\begin{equation}\label{sigmaxyR2}
\sigma_{xy}(f,g)=\beta \lim_{s\ra 0+}\int_0^\infty e^{-st} \langle\langle J(\delta f,t); J(\delta g)\rangle\rangle dt .
\end{equation}
Note that it still depends on the exact profile of the electric potential $g$ as well as on the choice of $f$. To get the electric Hall conductivity one needs to take the limit where 
$g$ is linear over a very large region in $y$. One also has to set $f(p)=\theta(x(p)-a)$ and average over $a$. We will see in the next section that at $T=0$ the precise choice of $f$ and $g$ becomes immaterial.

\subsection{Zero-temperature electric Hall conductance as a topological invariant}
\label{sec:invariance of sigma}

In this section we argue that for a gapped system at $T=0$ the electric Hall conductance  $\sigma_{xy}(f,g)$ is independent of the precise choice of functions $f$ and $g$ and unchanged under variations of the Hamiltonian which do not close the gap. This is an adaptation of the arguments of Niu and Thouless \cite{NiuThouless}.
We will also make use of the recent rigorous results on the decay of certain correlation functions in gapped systems obtained by H. Watanabe \cite{Watanabe}. 
Ref. \cite{Watanabe} assumes that the system is finite, so strictly speaking we need a generalization of these results to infinite systems. This generalization is straightforward, since Watanabe's estimates are uniform in the system's size.

After specializing to $T=0$, we follow Ref. \cite{NiuThouless} and rewrite $\sigma_{xy}(f,g)$ in terms of the many-body Green's function  $G=(z-H)^{-1}$:
\begin{equation}
\sigma_{xy}(f,g)=  -i\oint_{z=E_0} \frac{dz}{2\pi i} \text{Tr}\left( G J(\delta f) G^2 J(\delta g) \right).
\end{equation}
Here $E_0$ is the energy of the ground state, the contour of integration encloses the point $z=E_0$ counter-clockwise and trace is taken over the Hilbert space of the whole system. We also denote by $a$ the $x$-coordinate of the mid-line of the vertical strip where $J(\delta f)$ is supported, and denote by $b$ the $y$-coordinate of the mid-line of the horizontal strip where $J(\delta g)$ is supported.

First we will argue that shifting $f\mapsto f+f_0$, where $f_0(p)$ depends only on $x(p)$ and is compactly supported in the $x$-direction, does not affect $\sigma_{xy}(f,g)$. Under such a shift $\sigma_{xy}(f,g)$ changes by 
\begin{equation}
-i\oint_{z=E_0} \frac{dz}{2\pi i} \text{Tr}\left( G J(\delta f_0) G^2 J(\delta g) \right)=-\oint_{z=E_0} \frac{dz}{2\pi i} \text{Tr}\left( G [H,Q(f_0)] G^2 J(\delta g) \right).
\end{equation}
Using the identity $[H,A]=-[G^{-1},A]$, this expression can be written as
\begin{equation}\label{twoterms}
\oint_{z=E_0} \frac{dz}{2\pi i} \text{Tr}\left( Q(f_0) G^2 J(\delta g) \right)-\oint_{z=E_0} \frac{dz}{2\pi i} \text{Tr}\left( G Q(f_0) G J(\delta g) \right).
\end{equation}
The first term can be written as 
\begin{equation}
\oint_{z=E_0} \frac{dz}{2\pi i} \frac{\partial}{\partial z}\text{Tr}\left( Q(f_0) G J(\delta g) \right)
\end{equation}
and thus vanishes. The second terms is well-defined because according to \cite{Watanabe} correlators of the form
\begin{equation}\label{watanabestatic}
\oint_{z=E_0} \text{Tr}\left( G A G B\right)
\end{equation}
are exponentially small when the supports of $A$ and $B$ are separated by a large distance, and $Q(f_0)$ and $J(\delta g)$ are sums of local operators supported in a vertical and a horizontal strip, respectively. 

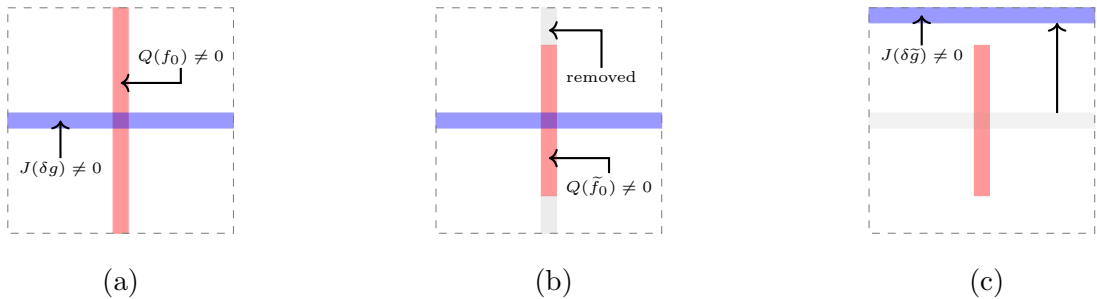
\begin{figure}[ht]
    \centering
% Part (a)
  \begin{subfigure}[b]{0.3\textwidth}
    \centering
    \begin{tikzpicture}
    % red line
    \filldraw[draw=red,color=red, opacity=0.4] (-0.1,-1.5) rectangle (0.1,1.5);
    % blue line
    \filldraw[draw=blue,color=blue, opacity=0.4] (-1.5,-0.1) rectangle (1.5,0.1);
    % box
    \draw[-,color=gray,dashed] (-1.5,1.5) -- (1.5,1.5)--(1.5,-1.5) -- (-1.5,-1.5)--(-1.5,1.5);
    % label
     \draw[black,thick,<-] (0,0.5) -- (0.8,0.5) -- (0.8,0.7) node[above=-0.1 cm,align=left]{\tiny $Q(f_0)\ne 0$};
     \draw[black,thick,<-] (-0.8,0) -- (-0.8,-0.5) node[below=-0.1 cm,align=left]{\tiny $J(\delta g)\ne 0$};
    \end{tikzpicture}
    \caption{}
\end{subfigure}
\hfill
% Part (b)
\begin{subfigure}[b]{0.3\textwidth}
    \centering
    \begin{tikzpicture}
    % red line
    \filldraw[draw=red,color=red, opacity=0.4] (-0.1,-1) rectangle (0.1,1);
    % blue line
    \filldraw[draw=blue,color=blue, opacity=0.4] (-1.5,-0.1) rectangle (1.5,0.1);
    % gray line
    \filldraw[draw=gray,color=gray, opacity=0.15] (-0.1,1) rectangle (0.1,1.5);
    \filldraw[draw=gray,color=gray, opacity=0.15] (-0.1,-1) rectangle (0.1,-1.5);
    % box
    \draw[-,color=gray,dashed] (-1.5,1.5) -- (1.5,1.5)--(1.5,-1.5) -- (-1.5,-1.5)--(-1.5,1.5);
    % labels
     \draw[black,thick,<-] (0,1.2) -- (0.7,1.2) -- (0.7,0.7) node[below=-0.1cm,align=left]{\tiny removed};
     \draw[black,thick,<-] (0,-0.5) -- (0.8,-0.5) -- (0.8,-0.7) node[below=-0.1 cm,align=left]{\tiny $ Q(\widetilde f_0)\ne 0$};
    \end{tikzpicture}
     \label{bluered}
    \caption{}
\end{subfigure}
% Part (c)
\hfill
\begin{subfigure}[b]{0.3\textwidth}
    \centering
    \begin{tikzpicture}
    % red line
    \filldraw[draw=red,color=red, opacity=0.4] (-0.1,-1) rectangle (0.1,1);
    % blue line
    \filldraw[draw=blue,color=blue, opacity=0.4] (-1.5,1.3) rectangle (1.5,1.5);
    % blue line
    \filldraw[draw=gray,color=gray, opacity=0.1] (-1.5,-0.1) rectangle (1.5,0.1);
    % box
    \draw[-,color=gray,dashed] (-1.5,1.5) -- (1.5,1.5)--(1.5,-1.5) -- (-1.5,-1.5)--(-1.5,1.5);
    % arrow
    \draw[->,thick] (1,0.1) -- (1,1.3);
    % labels
    \draw[black,thick,<-] (-0.8,1.4) -- (-0.8,1) node[below=-0.1 cm,align=left]{\tiny $J(\delta \widetilde g)\ne 0$};
    \end{tikzpicture}
     \label{bluered2}
    \caption{}
\end{subfigure}
    \caption{\footnotesize (a) The red vertical line represents the  support of $Q(f_0)$, the blue horizontal line represents the support of $J(\delta g )$.  (b) Grey parts are far away from the blue line, give a negligible contribution and can be dropped. (c) One can use the conservation law to move the blue line so that the blue and red lines are separated by a large distance. }
    \label{cycle deformation fig}
\end{figure}

As a matter of fact, the second term in (\ref{twoterms}) is also zero. To see this, let us replace the function $f_0$ with a function $\tilde f_0$ which is equal to $f_0$ for $|y-b|<L/2$ but vanishes for $|y-b|\geq L/2$. If $L$ is large, the exponential decay of the correlator (\ref{watanabestatic}) implies that the second term in (\ref{twoterms}) changes by an amount of order $L^{-\infty}$. Let $\tilde g$ denote the translate of $g$ in the $y$ direction by $L$, see Fig \ref{cycle deformation fig}. Clearly, $g_0=g-\tilde g$ is a function which depends only on $y(p)$ and is compactly supported in the $y$ direction. Therefore 
\begin{multline}
\oint \frac{dz}{2\pi i} \text{Tr}\left( G Q(f_0) G J(\delta g) \right)=\\
=\oint \frac{dz}{2\pi i} \text{Tr}\left( G Q(\widetilde f_0) G J(\delta\tilde g) \right)-i\oint\frac{dz}{2\pi i} \text{Tr}\left( G Q(\tilde f_0) G [H,Q(g_0)] \right)+O(L^{-\infty}).
\end{multline}
The first term here is of order $L^{-\infty}$ since the supports of $Q(\tilde f_0)$ and $J(\delta\tilde g)$ are separated by a distance of order $L$. The second term is zero, since
\begin{multline}
\oint\frac{dz}{2\pi i} \text{Tr}\left( G Q(\tilde f_0) G [H,Q(g_0)] \right)=-\oint\frac{dz}{2\pi i} \text{Tr}\left( G Q(\tilde f_0) G [G^{-1},Q(g_0)]\right)=\\
=-\oint\frac{dz}{2\pi i} \text{Tr}\left( G [Q(\tilde f_0),Q(g_0)]\right)=0,
\end{multline}
due to ultra-locality of the charge. Taking the limit $L\ra\infty$, we conclude that the second term in (\ref{twoterms}) is zero. This concludes the proof that $\sigma_{xy}(f,g)$ is independent of the precise choice of $f$. Independence of $g$ is proved similarly.

Note that the status of $f$ and $g$ was somewhat different until now. The function $g$ describes the profile of the electric potential and thus is a smeared step-function of nonzero width. The physically preferred value for  $f$ was an unsmeared  step-function of the $x$ coordinate. However, the difference between a smeared and unsmeared step-function is a function $f_0$ supported on an interval. The above argument shows that for $T=0$ shifting $f\mapsto f+f_0$ does not affect $\sigma_{xy}$. Thus at $T=0$ we can take both $f$ and $g$ to be unsmeared step-functions centered at $x=a$ and $y=b$, respectively. Exchanging $x$ and $y$ is then the same as exchanging $a$ and $b$. It is easy to see that $\sigma_{xy}$ is anti-symmetric under such an exchange, hence at $T=0$ it coincides with $\sigma^A$. This is to be expected, since at $T=0$ the dissipative part of the conductance tensor vanishes. 

Next we show that deformations of the Hamiltonian which do not close the energy gap do not affect $\sigma_{xy}(f,g)$. It is sufficient to show this for families of Hamiltonians of the form $H(\lambda)=H+\lambda V$, where $V$ is a local operator supported on a region of a fixed diameter $D$. The general case is an immediate consequence, since we can write an arbitrary deformation as a sum of such deformations. 

As explained above, we can choose $f$ and $g$ to be step-functions centered at $x=a$ and $y=b$, respectively. We will denote the corresponding current operators $J_a$ and $J_b$ and write
\begin{equation}\label{localized conductivity}
\sigma^A=  -i\oint_{z=E_0} \frac{dz}{2\pi i} \text{Tr}\left( G J_a G^2 J_b \right). 
\end{equation}
Since changing $a$ and $b$ does not affect $\sigma^A$, we can choose them so that the distance between the support of the perturbation $V$ and the lines $x=a$ and $y=b$ is of order $L$ where $L$ is arbitrarily large.
The variation of (\ref{localized conductivity}) under the deformation of the Hamiltonian is proportional to
\begin{align}
\begin{split}
\frac{\partial }{\partial \lambda}\oint_{z=E_0} \frac{dz}{2\pi i} \text{Tr}\left(G J_a G^2 J_b\right) =  \oint_{z=E_0} \frac {dz}{2\pi i}\left\{{\rm Tr \,} \left(GVG J_a G G  J_b\right)\right.\\
 \left.+{\rm Tr \,} \left(G J_a G VG G J_b\right)+{\rm Tr \,} \left(G J_a G GVG J_b\right)  \right\} ,
\end{split}
\end{align}
where we have used the fact that variations of $J_a, J_b$ are zero since the supports of $J_a$ and $J_b$ are more than a distance $2\R$ away from the support of $V$. We also used  
\begin{align}
\frac{\partial G}{\partial \lambda} &= G \frac{\partial H}{\partial \lambda} G =  GVG.
\end{align}
Subtracting a total derivative
\begin{align}
\begin{split}
0&= \oint \frac {dz}{2\pi i} \,{\rm Tr \,} \frac{\partial}{\partial z} \left(G J_a G VG J_b\right) =  \oint \frac {dz}{2\pi i} \left\{ {\rm Tr \,}  \left(GG J_a G V G J_b\right) \right. \\  &\qquad\qquad\qquad \left.+{\rm Tr \,}  \left(G J_a GG V G J_b\right)+{\rm Tr \,}  \left(G J_a G V GG J_b\right) \right\},
\end{split}
\end{align}
from the above expression, we get
\begin{align}
\begin{split}
\frac{\partial }{\partial \lambda}\oint_{z=E_0} \frac{dz}{2\pi i} \text{Tr}\left(G J_a G^2 J_b\right)=- \oint \frac {dz}{2\pi i} {\rm Tr \,} \left( \left[ V, G  J_a G\right] G J_b G\right).
\end{split}
\end{align}
In Appendix \ref{appendix: watanabe} we show that correlators of the form
\begin{align}
\begin{split}
  \oint \frac {dz}{2\pi i} {\rm Tr \,} \left( \left[A, G  BG\right] G CG\right),
\end{split}
\end{align}
where $A,B,C$ are local operators and the support of $A$ is away from the support of $B$, are exponentially suppressed for gapped systems. Therefore we have
\begin{align}
\begin{split}
\frac{\partial \sigma^A}{\partial \lambda} =-i\frac{\partial }{\partial \lambda}\oint_{z=E_0} \frac{dz}{2\pi i} \text{Tr}\left(G J_a(x_0) G^2 J_b(y_0)\right)= O(L^{-\infty}).
\end{split}
\end{align}
Since $L$ can be made arbitrarily large, this concludes the proof.

\section{Thermal Hall Conductance}

\subsection{Energy currents and energy magnetization on a lattice}

For a quantum system on a lattice $\Lambda\subset\RR^d$, the energy current from site $q$ to site $p$ is an operator $J^E_{pq}$ which satisfies
\begin{equation}\label{energy_conservation_equation}
\frac{dH_q}{dt}=-\sum_{p\in\Lambda} J^E_{pq}.
\end{equation}
An obvious solution is \cite{Mahan,Kitaev}
\begin{equation}\label{energycurrent}
J^E_{pq}=-i[H_p,H_q].
\end{equation}
Since $[H_p,H_q]=0$ whenever $|p-q|>2\R$, $J^E_{pq}$ is nonzero only when $p,q$ are nearby. The energy current from $B$ to $A=\Lambda\backslash B$ is defined to be
\begin{equation}
J^E(A,B)=\sum_{p\in A} \sum_{q\in B} J^E_{pq}=J^E(\delta\chi_B),
\end{equation}
where $\chi_B$ is the same as before. 

As in the case of the electric current, the expression for $J^E_{pq}$ is not unique. One can always make a replacement 
\begin{equation}\label{energy_current_ambiguity}
J^E_{pq}\mapsto J^E_{pq}+\sum_r U^E_{pqr},
\end{equation} 
where the operator $U^E_{pqr}$ is skew-symmetric under the exchange of the points $p,q,r$ and vanishes if any two of them are farther than some fixed distance. The modified energy current is physically equivalent to $J^E_{pq}$. Physical quantities, such as $J^E(A,B)$, are not affected by such modifications. It is shown in Appendix \ref{appendix:math} that this is the only ambiguity in the definition of the energy current, therefore the expression (\ref{energycurrent}) is essentially unique. In contrast, there is no simple and general expression for the energy current in continuum systems. This is one of the reasons we prefer to study lattice systems.

From now on we again specialize to 2d lattice systems.
In an equilibrium state we have
\begin{equation}\label{JEclosed}
\sum_p \langle J^E_{pq} \rangle=-\left\langle \frac{dH_q}{dt}\right\rangle=0.
\end{equation}
This suggests that there might exist a function $M^E:\Lambda\times\Lambda\times\Lambda\ra \RR$ which is skew-symmetric under the exchange of the arguments, decays rapidly when any two of the arguments are far apart, and satisfies
\begin{equation}\label{energymagnetization}
\langle J^E_{pq}\rangle=\sum_r M^E_{pqr}.
\end{equation}
It is easy to see that this expression automatically satisfies Eq. (\ref{JEclosed}). The decay property is required to make the sum over $r$ convergent. It is shown in Appendix \ref{appendix:math} that such an $M^E$ always exists. 

The equation (\ref{energymagnetization}) is a lattice analog of the continuum equation
\begin{equation}\label{continuumenergymagnetization}
\langle J^E_k(\br)\rangle=-\eps_{kj}\partial_j M^E(\br),
\end{equation}
which defines the ``energy magnetization'' $M^E(\br)$ of a 2d system \cite{Cooperetal,Niuetal}. Thus $M^E_{pqr}$ is a lattice analog of energy magnetization. Note that in the continuum energy magnetization is a function of spatial coordinates, while on the lattice it is a functions of three points. 

Unfortunately, there is no preferred choice of $M^E$, either in the continuum or on the lattice. This is more apparent in the continuum, where it is obvious that shifting $M^E(\br)\mapsto M^E(\br)+m^E$, where $m^E$ is independent of coordinates but may depend on parameters of the system, leaves $\langle J^E\rangle$ invariant. But the reason is essentially topological, and the ambiguity is present on a lattice as well. This lattice ambiguity {\it is  not} the obvious freedom to make a redefinition 
\begin{equation}\label{energy magnetization ambiguity}
M^E_{pqr}\mapsto M^E_{pqr}-\sum_s N_{pqrs},
\end{equation}
where $N_{pqrs}$ is a real-valued function of four points which decays rapidly when any of two of the points are far apart. The ambiguity (\ref{energy magnetization ambiguity}) is analogous to (although distinct from) the ambiguity (\ref{energy_current_ambiguity}) in the definition of the energy current and is harmless. For example, it will be shown below that it does not affect the thermal Hall energy flux. But for 2d systems there is a further ambiguity. For simplicity, let us take $\Lambda$ to be a regular triangular lattice in $\RR^2$. Then an obvious solution to the equation $\sum_r m^E_{pqr}=0$ is to take $m^E_{pqr}=\pm 1$ for any three points which are vertices of an elementary triangle of $\Lambda$ and zero in all other cases. The sign is determined by the orientation of the triangle $[pqr]$ relative to the orientation of $\RR^2$. The freedom to add to $M^E_{pqr}$ a multiple of $m^E_{pqr}$ is analogous to the freedom to add a constant $m^E$ to the continuum energy magnetization $M^E(\br)$. 

One can partially fix the ambiguity by requiring $M^E$ to be a local quantity. In the continuum case, this means that $M^E(\br)$ cannot depend on the values of the parameters of the Hamiltonian far from $\br$. This leaves the freedom to shift $M^E(\br)$ by $m^E(\br)$ which satisfies
\begin{equation}
\partial_j m^E(\br)=0,\quad \lim_{|\br-\br'|\ra\infty}\frac{\delta m^E(\br)}{\delta \lambda(\br')}=0,
\end{equation}
where $\lambda(\br)$ is the value of a parameter $\lambda$ at a point $\br'$. This implies that $m^E(\br)$ depends neither on $\br$ nor on $\lambda$.  Still, shifting $M^E(\br)$ by a constant independent of any parameters is allowed. This shows that $M^E(\br)$ is not a physical quantity. The situation on the lattice is similar.

On the other hand, the above arguments show that derivatives of $M^E(\br)$ with respect to parameters of the system are free from ambiguities. Indeed, for a special class of continuum Hamiltonians, Ref.~\cite{Niuetal} derived a well-defined microscopic formula for the derivative of the volume-average of $M^E(\br)$ with respect to the chemical potential. It was noticed by A. Kitaev \cite{Kitaev} that on the lattice the situation is even better: there is a natural a formula for the derivatives of $M^E_{pqr}$ with respect to arbitrary variations of the Hamiltonian. That is, if the Hamiltonian depends on some parameters $\lambda^\ell$, then there is a manifestly local solution of the system of equations
\begin{equation}\label{energymagnetizationderivative}
\frac{\partial}{\partial\lambda^\ell}\langle J^E_{pq}\rangle=\sum_r \mu^E_{pqr,\ell}.
\end{equation}
If we assemble the quantities $\mu^E_{pqr,\ell}$ into a 1-form $\mu^E_{pqr}=\mu^E_{pqr,\ell} d\lambda^\ell$ on the parameter space, then the above equation is solved by
\begin{equation}\label{mue}
\mu^E_{pqr}=-\beta\langle\langle dH_p; J^E_{qr}\rangle\rangle-\beta\langle\langle dH_r; J^E_{pq}\rangle\rangle-\beta\langle\langle dH_q; J^E_{rp}\rangle\rangle.
\end{equation}
Here $d = \sum_{l}^{} d\lambda^\ell \frac{\partial }{\partial \lambda^\ell}$ is the exterior derivative on space of local Hamiltonians. 
The identity (\ref{energymagnetizationderivative}) is easily verified using properties of the Kubo pairing (see Appendix \ref{appendix:Kubo pairing}) and the definition of the energy current (\ref{energycurrent}).

The quantity $\mu^E_{pqr,\ell}$ has the meaning of the derivative of the energy magnetization with respect to $\lambda^\ell$. If the correlation length is finite, $\mu^E_{pqr}$ is exponentially small when any two of the points $p,q,r$ are far from each other. It is also apparent that $\mu^E_{pqr,\ell}$ does not depend on the state of the system far from the points $p,q,r$. In other words, it is a local quantity.

The expression $\mu^E_{pqr}$ is not unique: one can always make a replacement 
\begin{equation}\label{mueambiguity}
\mu^E_{pqr}\mapsto \mu^E_{pqr}-\sum_s \nu^E_{pqrs},
\end{equation}
where $\nu^E_{pqrs}$ is a 1-form on the parameter space which depends on four lattice points $p,q,r,s$, is skew-symmetric under the exchange of these points, and decays rapidly when any two of the points are far apart. However, as we will see below, this ambiguity does not affect physical quantities which we compute. 

\subsection{Kubo formula for the derivatives of the thermal Hall conductance}\label{sec:thermal Hall conductance}

To derive a Kubo formula for derivatives of the thermal Hall conductance we follow the same strategy as in the case of electric Hall conductance. Following Luttinger \cite{Luttinger}, we perturb the Hamiltonian by a term
\begin{equation}
\Delta H(t)=\eps e^{st}  \sum_{p\in\Lambda} g(p) H_p,\quad t\in (-\infty,0].
\end{equation}
It is shown in \cite{Luttinger} that this is equivalent to a time-dependent and space-dependent infinitesimal temperature deformation
$$
\delta T(t,{\bf r})=\varepsilon T e^{st} g({\bf r}). 
$$
As in the electric case, we cannot take $g$ to be a smeared step-function of $y$, since then the change in the expectation value of the net energy current across a line $x=a$ will be ill-defined. Instead we take $g$ to be a function as in Fig. \ref{g hat fig}, and consider an inhomogeneous system whose Hamiltonian depends on a parameter $\lambda$ which varies with $y$ as in Fig. \ref{g hat fig}. This allows one to compute the derivative of the thermal Hall conductance with respect to $\lambda$. 

One difference compared to the electric case is that the energy current operator $A=J^E(\delta f)$ now has an explicit dependence on $\eps$ (the magnitude of the perturbation). This happens because $[H_p,H_q]\neq 0$, in general. The change in $A$ due to this explicit dependence is 
\begin{equation}
\Delta A=-\frac \eps 2 \sum_{p,q} i [H_p,H_q](f(q)-f(p)) (g(p)+g(q)).
\end{equation}
The corresponding change in the expectation value of $A$ is
\begin{equation}\label{explicit change}
\frac \eps 2 \sum_{p,q}\langle J^E_{pq} \rangle (f(q)-f(p))(g(p)+g(q))=\frac \eps 3 \sum_{p,q,r}M^E_{pqr}(f(q)-f(p))(g(q)-g(r)).
\end{equation}
Here we used skew-symmetry of $M^E_{pqr}$ with respect to arbitrary permutations of $p,q,r$. 

Since $M^E_{pqr}$ decays rapidly when $q$ and $r$ are far apart, and since $g(q)-g(r)$ vanishes when $q$ and $r$ are both in a region where $g$ is constant, eq. (\ref{explicit change}) receives contributions only from the regions $I$ and $II$ where the temperature gradient is nonzero.
We make this explicit by writing $\delta g=\delta g_I+\delta g_{II}$, where $g_{II}(p)$ depends only on $y(p)$ and interpolates between $0$ and $1$ as one moves from $y=+\infty$ to the intermediate region, and $g_I(p)$ depends only on $y(p)$ and interpolates between $1$ and $0$ as one moves from the intermediate region to $y=-\infty$. If the temperature gradients in regions $I$ and $II$ are  equal and opposite, $\delta g_I$ is minus the translate of $\delta g_{II}$. In these two regions the parameter $\lambda$ takes constant values $\lambda_1$ and $\lambda_2$, respectively.  Therefore the expression (\ref{explicit change}) can be written as
\begin{equation}\label{explicit change 2}
2\eps (\lambda_2-\lambda_1)\,\mu^E(\delta g_{II}\cup\delta f)+O\Big((\lambda_2-\lambda_1)^2\Big),
\end{equation}
where $\mu^E_{pqr}=\partial M^E_{pqr}/\partial\lambda$, and we introduced a shorthand
\begin{equation}
\mu^E(\delta f_1\cup \delta f_2)=\frac 1 6\sum_{p,q,r} \mu^E_{pqr}(f_1(q)-f_1(p))(f_2(r)-f_2(q))
\end{equation}
for any two functions $f_1,f_2:\Lambda\ra\RR$. For generic functions $f_1,f_2$ the triple sum over $p,q,r$ has a large-volume divergence  which arises from the region where $p,q,r$ are all close together. However, for $f_1=g_{II}$ and $f_2=f$ it is easy to check that the summation is absolutely convergent, so the expression is well-defined. Using eq. (\ref{energymagnetizationderivative}) and the skew-symmetry of $\mu^E_{pqr}$ with respect to $p,q,r$ one can check that the $\mu^E(\delta f_1\cup\delta f_2)$ is unchanged under a redefinition (\ref{mueambiguity}) and is skew-symmetric under the exchange of $f_1$ and $f_2$. Such checks become routine if one uses the machinery of Appendix \ref{appendix:math}.

Combining (\ref{explicit change 2}) with the change in $\langle A\rangle$ arising from the change in the state of the system, we get
\begin{equation}
\Delta\langle A\rangle\approx\eps (\lambda_2-\lambda_1)\frac{\partial}{\partial\lambda}\left[\beta\lim_{s\ra 0+}\int_0^\infty e^{-st} \langle\langle J^E(\delta f,t); J^E(\delta g)\rangle\rangle dt\right]+2\eps(\lambda_2-\lambda_1)\mu^E(\delta g\cup\delta f).
\end{equation}
Here to simplify notation we denoted by $\delta g$ what previously was denoted $\delta g_{II}$. That is, $g(p)$ now denotes a function of $y(p)$ which interpolates from $1$ at $y\ll 0$ to $0$ at $y\gg 0$. 
On the other hand, the expected net energy current across the line $x=a$ is
$$
-\eps T \int_{-\infty}^{+\infty} \kappa_{xy} \partial_y g\, dy=\eps T\int_{\lambda_1}^{\lambda_2} \frac{\partial\kappa_{xy}}{\partial\lambda} d\lambda .
$$
Comparing these two expressions we get a formula for the $\lambda$-derivative of the thermal Hall conductance:
\begin{equation}\label{derivative of thermal Hall}
d\kappa_{xy}(f,g)=d\left[\beta^2\lim_{s\ra 0+}\int_0^\infty e^{-st} \langle\langle J^E(\delta f,t); J^E(\delta g)\rangle\rangle dt \right]-2\beta\mu^E(\delta f\cup\delta g).
\end{equation}

Unlike in the electric case, there is no canonical formula for $\kappa_{xy}(f,g)$. We can still define the difference of thermal Hall conductances of two materials $\ma$ and $\mb$ by integrating the 1-form $d\kappa_{xy}$ along a path in the parameter space connecting $\ma$ and $\mb$. This path must avoid phase transitions, otherwise objects like $\mu^E(\delta f\cup\delta g)$ might diverge. 

\subsection{Path-independence of the thermal Hall conductance}
\label{path independence section}

We have defined a 1-form $d\kappa_{xy}$ on the space of parameters of a lattice system whose integral along a curve $\Gamma$ can be identified with the difference of thermal Hall conductances of the initial and final points of $\Gamma$. The definition of the 1-form depended on the rapid spatial decay of the Kubo pairings of local operators. Thus when choosing a curve connecting two points $\ma$ and $\mb$ in the parameter space, one needs to avoid loci where phase transitions occur. Since we are allowed to enlarge the parameter space by adding arbitrary local terms to the Hamiltonian, it is very plausible that such a curve exists for any two points $\ma$ and $\mb$. Indeed, phase transitions at nonzero temperatures are usually associated with spontaneous symmetry breaking and typically can be turned into cross-overs by adding suitable symmetry-breaking perturbations. Quantum phase transitions at zero-temperature defy the symmetry-breaking paradigm, but as explained in the next section temperature can be considered as one of the parameters of the Hamiltonian, and at non-zero temperature quantum phase transitions become cross-overs.

An important consistency requirement is that the difference of the thermal Hall conductances thus computed does not depend on the choice of $\Gamma$. To show this, consider an  arbitrary closed loop $\Gamma$ in the parameter space. By  assumption, the ``Kubo'' part of the thermal Hall conductance 
\begin{equation}
    \kappa_{xy}^{\rm Kubo}(f,g)=\beta^2\lim_{s\ra 0+}\int_0^\infty e^{-st} \langle\langle J^E(\delta f,t); J^E(\delta g)\rangle\rangle dt ,
\end{equation}
is well-defined for each point of $\Gamma$. Therefore  $d\kappa^{\rm Kubo}_{xy}(f,g)$ is an exact 1-form and its integral over any closed curve vanishes. 

We are going to argue that the energy  magnetization contribution $\mu^E(\delta f\cup\delta g)$ is also exact. This is a 1-form on the parameter space which depends on $f$ and $g$. Its physical meaning is the differential of the energy magnetization in the region where both $f$ and $g$ vary substantially. We would like to show that the integral of this 1-form along any loop $\Gamma$ avoiding phase transitions is zero. Heuristically, this must be true in order to avoid contradiction with the theorem about the absence of net energy currents in equilibrium quasi-1d systems \cite{energyBloch}. Imagine slowly varying the parameters of the system as a function of $y\in[0,L]$ while following a loop  $\Gamma$. Then we can  compactify the $y$ direction with period $L$, and regard this as a quasi-1d system. If $L$ is large compared to the correlation length, this should not affect local properties, including the differential of the energy magnetization $\mu^E$. The energy current in the $x$ direction can be computed using the continuum equation (\ref{continuumenergymagnetization}). Since the net energy current should vanish, we 
get
\begin{equation}
0=\int \langle J^E_x\rangle dy=\int_0^L \partial_y M^E dy\simeq \int \partial_\lambda M^E d\lambda =\int_\Gamma \mu^E.
\end{equation}
The error in this computation should become arbitrarily small for $L\ra\infty$, so we get the desired result. A more precise argument is given in Appendix \ref{appendix:path}.

\subsection{A relative invariant of gapped 2d lattice systems}
\label{sec:thermalHalltopinvariance}

In this section we use the 1-form $d\kappa_{xy}$ to define a relative topological invariant of gapped 2d lattice systems at zero temperature. We anticipate that in the case when both lattice systems admit a conformally-invariant edge, the invariant will be equal to $\pi/6$ times the difference of the chiral central charges for the two systems. We cannot necessarily connect two such systems by a curve $\Gamma$ in the space of Hamiltonians without encountering a bulk phase transition. If we could, this would mean that they are in the same zero-temperature phase, and then by the result of \cite{energyBloch} they would have to have the same chiral central charge for the edge modes, and therefore the relative invariant would vanish. Rather, the idea is to treat the temperature $T$ as yet another parameter, and connect the two systems in the enlarged parameter space $(T,\lambda).$ At positive temperatures quantum phase transitions are smoothed out into cross-overs, and the two systems can now be deformed into each other while maintaining a finite correlation length.

Formally, the temperature can be regarded as a parameter because re-scaling the temperature by a positive factor is equivalent to re-scaling the Hamiltonian by the inverse factor. Therefore one can extend the form $\kappa_{xy}(f,g)$ to the open subset of the enlarged parameter space given by $T>0$. In detail, this is done as follows. Given a Hamiltonian $H$, we define a one-parameter family of Hamiltonians by  $H(\lambda^0)=\lambda^0 H$ (the Hamiltonian $H$ still depends on other parameters $\lambda^\ell$ which we collectively call $\lambda$ opposed to specific overall scaling parameter $\lambda^0$). Then the above mentioned scaling symmetry implies
\begin{equation}\label{homogeneity}
T\frac{d}{dT}\frac{\kappa_{xy}^{\rm Kubo}(f,g)}{T}=-\left.\lambda^0\frac{d}{d\lambda^0}\right|_{\lambda^0=1} \frac{\kappa_{xy}^{\rm Kubo}(f,g;\lambda^0)}{T},
\end{equation}
where $\kappa_{xy}^{\rm Kubo}(f,g;\lambda^0)$ denotes the Kubo part of $\kappa_{xy}$ computed with the Hamiltonian $H(\lambda^0)$. We have to divide $\kappa_{xy}$ by $T$ in order to get an observable which is invariant under the rescaling $H\mapsto \lambda^0 H, T\mapsto\lambda^0 T$. Similarly, we have
\begin{equation}
2T\frac{d}{dT}\left(\beta^2 M^E_{pqr}\right)=-2 \left.\lambda^0\frac{d}{d\lambda^0}\right|_{\lambda^0=1}\left(\beta^2 M^E_{pqr}\right)=\frac{2}{T^2}\tau^E_{pqr},
\end{equation}
where $\tau^E_{pqr}$ is $-\mu^E_{pqr}$ with $dH_p$ replaced with $H_p$:
\begin{equation}
\tau^E_{pqr}=\beta\langle\langle H_p; J^E_{qr}\rangle\rangle+\beta\langle\langle H_r; J^E_{pq}\rangle\rangle+\beta\langle\langle H_q; J^E_{rp}\rangle\rangle.
\end{equation}

We can now define a 1-form on the subset $T>0$ of the enlarged parameter space which represents the total derivative of $\kappa_{xy}(f,g)/T$:
\begin{equation}\label{Psi}
\Psi(f,g)=\frac{d\kappa^{\rm Kubo}_{xy}(f,g)}{T}-\frac{2}{T^2}\mu^E(\delta f \cup\delta g)+\frac{d}{dT}\left(\frac{\kappa^{\rm Kubo}_{xy}(f,g)}{T}\right) dT-2\tau^E(\delta f \cup\delta g) \frac{dT}{T^3} .
\end{equation}
Its integral around any closed curve in the $(T,\lambda)$ space is zero by the same argument as before, therefore $\Psi$ is exact.

Given any two gapped zero-temperature lattice systems $\ma$ and $\mb$, we would like to define a relative topological invariant by integrating $\Psi$ along a curve in the enlarged parameter space which connects $\ma$ and $\mb$. See Fig. 3b. We need to check three things: that the integral converges, that it does not change as one deforms $\ma$ and $\mb$ while keeping $T=0$ and finite correlation length, and that result of integration does not change as we modify the functions $f,g$ while keeping their asymptotic behavior fixed. Neither of these is obvious.  The $T$-component of the 1-form $\Psi$ is
\begin{multline}\label{Psi n}
\Psi_n(f,g)=\frac{d}{dT}\left(\frac{\kappa^{\rm Kubo}_{xy}( f, g)}{T}\right)-\frac{2}{T^3}{\tau^E}(\delta f\cup\delta g)\\
=-\frac{1}{T^3}\left[\left.\frac{d}{d\lambda^0}\right|_{\lambda^0=1}\int_0^\infty \beta\langle\langle J^E_{\lambda^0}(\delta f,t);J^E_{\lambda^0} (\delta g)\rangle\rangle_{\lambda^0} dt+2\tau^E(\delta f\cup\delta g)\right].
\end{multline}
Here $\langle\langle\ldots\rangle\rangle_{\lambda^0}$ denotes the Kubo pairing at temperature $T$ with respect to the Hamiltonian $H(\lambda^0)=\lambda^0 H$, and $J^E_{\lambda^0}$ is the energy current for the Hamiltonian $H(\lambda^0).$ We denoted the $T$-component $\Psi_n$ to emphasize that it is the normal component to the boundary $T=0$ of the enlarged parameter space. The convergence of the integral of $\Psi$ requires the expression in parentheses to vanish faster than $T^2$ as $T\ra 0$. Similarly, the independence of the integral of $\Psi$ on the deformation of the endpoints requires the tangential component of $\Psi$,
\begin{equation}\label{Psi t}
\Psi_t(f,g)=\frac{1}{T^2}\left(d\int_0^\infty \beta\langle\langle J^E(\delta f,t);J^E(\delta g)\rangle\rangle dt-2\mu^E(\delta f\cup\delta g)\right).
\end{equation}
to vanish at $T=0$. Thus the expression in parentheses should vanish faster than $T^2$ as $T\ra 0$. 

In Appendix \ref{psi appednix} we argue (not completely rigorously) that both expressions vanish exponentially fast as $T\ra 0$. To see why this is plausible, consider eq. (\ref{Psi t}) for definiteness and denote the expression in parentheses as $\Omega(T)$. It is a 1-form on the space of parameters of the Hamiltonian. The first term in $\Omega(T)$ is the exterior derivative of the same kind of current correlator which defines the electric Hall conductance, except that the electric current $J$ is replaced with the energy current $J^E$. The key point is that at $T=0$ this correlator is sensitive mainly to the state of the system in a compact region $S\subset\RR^2$ which is an intersection of the vertical strip corresponding to $f$ and the horizontal strip corresponding to $g$. The same argument as in Section \ref{sec:invariance of sigma} shows that at $T=0$ the derivative of this correlator with respect to a deformation of the Hamiltonian localized at a distance $L$ from $S$ is of order $L^{-\infty}$. The same is true for the second term, because of the assumed decay of Kubo pairings. Since the sum of the two terms does not change as one varies  $f$ and $g$, $L$ can be made arbitrarily large, and we conclude that $\Omega(0)=0$ when evaluated on any deformation of the Hamiltonian supported on a quadrant in $\RR^2$. Therefore $\Omega(0)=0$ identically. Further, in the presence of the energy gap one expects the low-temperature expansion to have a finite radius of convergence,  therefore $\Omega(T)-\Omega(0)$ is exponentially suppressed for low $T$ (it is this part of the argument which is not rigorous). Combining these statements, we show that integral converges and independent of deformations of $\ma$ and $\mb$ which do not cross phase transitions. In order to show that the value of integral is independent of the shift of cochains $f,g$ we can use the fact that one form $\Psi(f,g)$ is exact and integral is given by difference of antiderivatives of $\Psi$ at endpoints. The latter can be shown to vanish if either $f$ or $g$ is hat-shaped as in Fig. \ref{g hat fig}. For more details see Appendix \ref{psi appednix}.

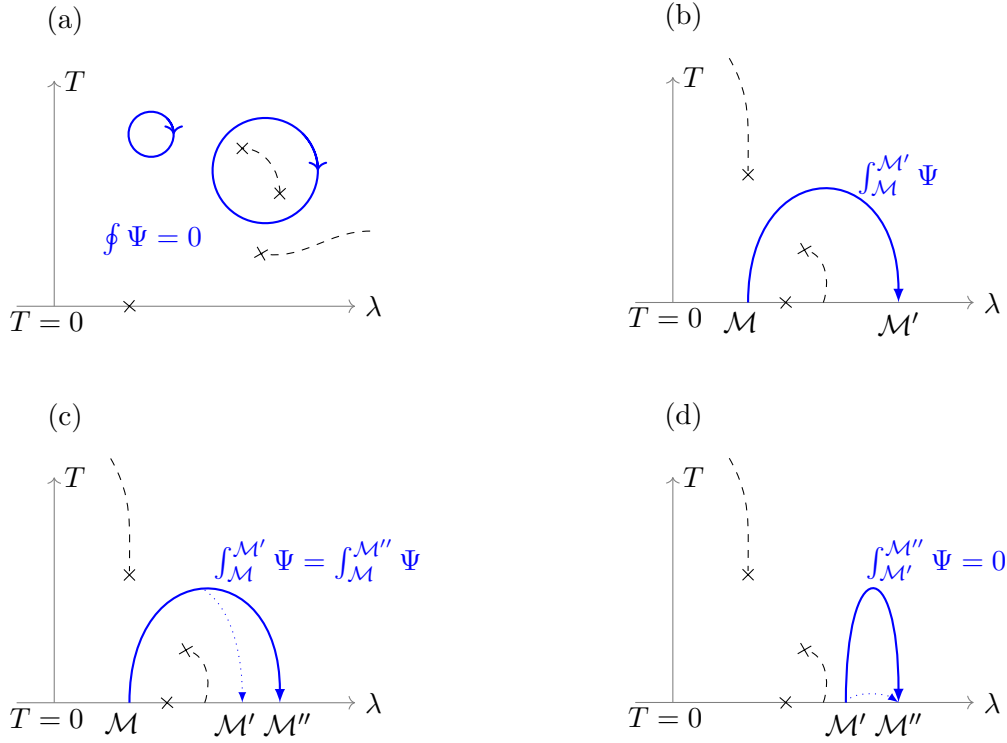
\begin{figure}[ht]
    \centering
% Part (a)
 \begin{subfigure}{0.3\textwidth}
    \caption{\qquad\qquad\qquad\qquad\qquad}
    \centering
   \begin{tikzpicture}
    % lambda axis
    \draw[->,color=gray] (-0.5,0) -- (4,0) node[right,color=black] {$\lambda$};
    % T axis
    \draw[->,color=gray] (0,0) -- (0,3) node[right,color=black] {$T$};
    % T=0 label 
    \draw (-0.1,-0.2) node {$T=0$}; 
    % phase transitions
    \draw (1,0) node[cross=2.5 pt,rotate=5] {};
    \draw[dashed] (4.2,1) edge[out=180,in=350] (2.75,0.7);
    \draw (2.75,0.7) node[cross=2.5 pt,rotate=75] {};
    \draw[dashed] (3,1.5) edge[out=90,in=0] (2.5,2.1);
    \draw   (3,1.5) node[cross=2.5 pt,rotate=-3] {};
    \draw   (2.5,2.1) node[cross=2.5 pt,rotate=0] {};
    % Integration countours
    \draw[thick, ->,blue] (3.3,2.3) arc (45:-360:0.7cm);% syntax (starting point coordinates) arc (starting angle:ending angle:radius)
    \draw[thick, ->,blue] (1.5,2.5) arc (45:-360:0.3cm);% syntax (starting point coordinates) arc (starting angle:ending angle:radius)
    \draw (1.3,0.9) node[blue] {$\oint \Psi=0$};
    \end{tikzpicture}
\end{subfigure}
\hfill
% Part (b)
  \begin{subfigure}{0.3\textwidth}
    \caption{\qquad\qquad\qquad\qquad\qquad}
    \centering
    \begin{tikzpicture}
    % lambda axis
    \draw[->,color=gray] (-0.5,0) -- (4,0) node[right,color=black] {$\lambda$};
    % T axis
    \draw[->,color=gray] (0,0) -- (0,3) node[right,color=black] {$T$};
    % T=0 label 
    \draw (-0.1,-0.2) node {$T=0$}; 
    % phase transitions
    \draw (1.5,0) node[cross=2.5 pt,rotate=5] {};
    \draw[dashed] (2,0) edge[out=70,in=350] (1.75,0.7);
    \draw (1.75,0.7) node[cross=2.5 pt,rotate=75] {};
    \draw[dashed] (0.75,3.25) edge[out=300,in=90] (1,1.75);
    \draw  (1,1.7) node[cross=2.5 pt,rotate=0] {};
    % Integration countour
    \draw[->, >=latex, thick,blue] (1,0)  to[controls=+(90:2) and +(90:2)] (3,0);
    \draw[black,below] (0.9,0) node[below,color=black] {$\ma$};
    \draw[black,below] (3,0) node[below,color=black] {$\mb$};
    \draw (3,1.7) node[blue] {$\int_{\ma}^\mb \Psi$};
    \end{tikzpicture}
\end{subfigure}
\hfill
\bigskip

    \centering
% Part (c)
  \begin{subfigure}{0.3\textwidth}
    \caption{\qquad\qquad\qquad\qquad\qquad}
    \centering
    \begin{tikzpicture}
    % lambda axis
    \draw[->,color=gray] (-0.5,0) -- (4,0) node[right,color=black] {$\lambda$};
    % T axis
    \draw[->,color=gray] (0,0) -- (0,3) node[right,color=black] {$T$};
    % T=0 label 
    \draw (-0.1,-0.2) node {$T=0$}; 
    % phase transitions
    \draw (1.5,0) node[cross=2.5 pt,rotate=5] {};
    \draw[dashed] (2,0) edge[out=70,in=350] (1.75,0.7);
    \draw (1.75,0.7) node[cross=2.5 pt,rotate=75] {};
    \draw[dashed] (0.75,3.25) edge[out=300,in=90] (1,1.75);
    \draw  (1,1.7) node[cross=2.5 pt,rotate=0] {};
    % Integration countours
    \draw[->, >=latex, thick,blue] (1,0) to[controls=+(90:2) and +(90:2)] (3,0);
    \draw[->, >=latex, dotted,blue] (2,1.5) to[controls=+(-35:0.5) and +(90:0.5)] (2.5,0);
    \draw[black,below] (0.9,0) node[below,color=black] {$\ma$};
    \draw[black,below] (2.4,0) node[below,color=black] {$\mb$};
    \draw[black,below] (3.1,0) node[below,color=black] {$\mc$};
    \draw (3.5,1.9) node[blue] {$\int_{\ma}^{\mb} \Psi=\int_{\ma}^{\mc} \Psi$};
    \end{tikzpicture}
\end{subfigure}
\hfill
% Part (d)
  \begin{subfigure}{0.3\textwidth}

    \caption{\qquad\qquad\qquad\qquad\qquad}
    \centering
    \begin{tikzpicture}
    % lambda axis
    \draw[->,color=gray] (-0.5,0) -- (4,0) node[right,color=black] {$\lambda$};
    % T axis
    \draw[->,color=gray] (0,0) -- (0,3) node[right,color=black] {$T$};
    % T=0 label 
    \draw (-0.1,-0.2) node {$T=0$}; 
    % phase transitions
    \draw (1.5,0) node[cross=2.5 pt,rotate=5] {};
    \draw[dashed] (2,0) edge[out=70,in=350] (1.75,0.7);
    \draw (1.75,0.7) node[cross=2.5 pt,rotate=75] {};
    \draw[dashed] (0.75,3.25) edge[out=300,in=90] (1,1.75);
    \draw  (1,1.7) node[cross=2.5 pt,rotate=0] {};
    % Integration countours
    \draw[->, >=latex, thick,blue] (2.3,0) node[below,color=black] {$\mb$} to[controls=+(90:2) and +(90:2)] (3,0) node[below,color=black] {$\mc$};
    \draw[->,>=latex, dotted,blue] (2.3,0) to[controls=+(60:0.2) and +(145:0.2)] (3,0);
    \draw (3.5,1.9) node[blue] {$\int_{\mb}^{\mc} \Psi=0$};
    \end{tikzpicture}
\end{subfigure}
\hfill
\bigskip
    \caption{ \footnotesize Phase diagrams. The horizontal axis represents a parameter of the Hamiltonian, the vertical axis is temperature. Dashed lines and crosses represent phase transitions. Blue lines are integration contours. (a) The integral of $\Psi$ along a loop is zero regardless of whether  there are phase transitions in the interior. (b) The invariant $I(\ma,\mb)$ for zero-temperature phases $\ma$ and $\mb$ can be computed by integrating $\Psi$ along the blue line. (c) The points $\mb$ and $\mc$ are in the same phase, therefore one expects the integrals along the solid and dotted blue lines to be the same. (d) The difference of the two paths can be deformed to a near-zero-temperature path from $\mc$ to $\mb$. $\Psi$ is exponentially small on this path. }
\end{figure}

There is another limit where one can evaluate $\Psi$, namely $T\ra\infty$. In this limit the expectation value $\langle A\rangle$ of a local operator $A$ becomes the normalized trace over the local Hilbert space, while the Kubo pairing becomes
\begin{equation}
\lim_{T\ra\infty}\langle\langle A;B\rangle\rangle=\langle AB\rangle-\langle A\rangle \langle B\rangle.
\end{equation}
Thus all components of $\Psi$ are of order $1/T^3$ for large $T$, and therefore the relative thermal Hall conductance of any two high-temperature states is of order $1/T^2$. Hence another natural choice of a reference state (apart from the trivial insulator at $T=0$) is the $T=\infty$ state. That is, one can define an absolute topological invariant of a gapped zero-temperature system $\ma$ by integrating the 1-form $\Psi$ along any path connecting $\ma$ to the $T=\infty$ state. 

The case of a Locally Commuting Projector Hamiltonian is particularly simple. In this case, since $J^E_{pq}=-i[H_p,H_q]=0$ for all $p,q$, the $T$-component of the 1-form $\Psi$ vanishes identically. Integrating $\Psi$ along a path $\Gamma$ along which only $T$ changes, we find that $\kappa^A(T)-\kappa^A(\infty)=0$. Thus the thermal Hall conductance relative to the $T=\infty$ state is zero for all temperatures.\footnote{Strictly speaking, to avoid potential phase transitions at $T>0$, one needs to work with a finite-volume version of $\Psi$ defined on torus. Its $T$-component still vanishes for a system described by a Local Commuting Projector Hamiltonian, so the integral from any $T$ to $T=\infty$ is still zero. Taking the infinite-volume limit we conclude that the relative thermal Hall conductance is identically zero.}  This implies that the chiral central charge of the edge modes must vanish for such a Hamiltonian. One can also show that the zero-temperature electric Hall conductance vanishes for such systems, but the proof is very different \cite{KapFid}. 

The case of gapped systems of free fermions is also fairly simple, since there are no phase transitions at any $T>0$, and one can again integrate $\Psi$ along a path with only $T$ varying. Then one only needs to know the $T$-component of $\Psi$, which can be evaluated in complete generality. This computation is performed in Appendix \ref{appednix:free fermions} where it is shown that
\begin{equation}\label{WF}
\int_{T=0}^{T=\infty} \Psi=\left. \frac{\kappa^A}{T}\right|_{T=\infty}-\left.\frac{\kappa^A}{T}\right|_{T=0}=-\frac{\pi^2}{3} \sigma^A,
\end{equation}
where $\sigma^A$ is the electric Hall conductance at $T=0$. If one defines $\kappa^A/T$ to vanish at $T=\infty$, then this can be regarded as a form of the Wiedemann-Franz law. Note however that it cannot be interpreted too naively. For example, since $\Psi$ is exponentially small for low $T$, most of the contribution to the integral (\ref{WF}) comes from $T$ of order of the energy gap. Although one can define the absolute thermal Hall conductance at temperature $T$ as 
\begin{equation}
\kappa^A(T)=T\int_{\infty}^T \Psi,
\end{equation}
and it will obey the Wiedemann-Franz law $\kappa^A\simeq \frac{\pi^2}{3} T \sigma^A$ at low $T$, $\kappa^A(T)$ is not determined by correlators measured at temperature $T$ and a fixed Hamiltonian.

\section{Concluding remarks}

We have derived a formula for the derivatives of the thermal Hall conductance with respect to parameters of the Hamiltonian and temperature. The relative thermal Hall conductance is obtained by integrating the derivative along a path in parameter space connecting the two materials. We have argued that this is the best one can do, since only differences of thermal Hall conductances of materials are well-defined physical quantities. What is usually measured in experiments is the thermal Hall conductance of a particular material relative to the vacuum. 

We also argued that for gapped 2d lattice systems the thermal Hall conductance at low $T$ is linear in $T$ up to exponentially small corrections. The slope of the thermal Hall conductance is a topological invariant, in the sense that it does not change under variations of the Hamiltonian which keep the correlation length finite. It can change only when the bulk undergoes a zero-temperature phase transition. 

This result can be interpreted as a form of bulk-boundary correspondence. Consider a strip of a gapped 2d material $\ma$ at temperatures below the gap as well as below any temperatures at which bulk phase transitions occur. Suppose there is an effective field theory description of this system which reproduces all observations. Since there are no bulk excitations, such an effective field theory describes only edge excitations. Let us assume that these edge excitations are described by a 1+1d CFT. There may also be  terms in the effective action which describe the bulk response to the external fields, such as the Chern-Simons term (if the system has a $U(1)$ symmetry and can be coupled to a background electromagnetic field) and the gravitational Chern-Simons term. However, such terms in the action do not contribute to the thermal Hall current at leading order in the temperature gradient \cite{Stone}. Thus the net energy current for the edge modes should be equal to $\kappa^A_{\ma{\mathcal M}_0}(T) \Delta T,$ where ${\mathcal M}_0$ is the vacuum and it is assumed that the temperature difference between the edges $\Delta T$ is much smaller than $T$. On the other hand, as explained in Section \ref{sec:intro}, the net energy current computed from CFT is equal to $\frac{\pi}{6} (c_R-c_L) T \Delta T$. Therefore the slope of $\kappa^A_{\ma{\mathcal M}_0}(T)$ at low $T$ is equal to $\frac{\pi}{6} (c_R-c_L)$.

\appendix

\section{Kubo canonical pairing} 
\label{appendix:Kubo pairing}

Kubo canonical pairing of two operators $A,B$ is defined as follows \cite{Kubo}:
\begin{equation}
\langle\langle A;B\rangle\rangle=\frac{1}{\beta}\int_0^\beta\langle A(-i\tau)B\rangle d\tau-\langle A\rangle\langle B\rangle.
\end{equation}
Here $\langle\ldots\rangle$ denotes average over a Gibbs state at temperature $T=1/\beta$ (or more generally, over a state satisfying the Kubo-Martin-Schwinger condition), and $A(-i\tau)=e^{H\tau}A e^{-H\tau}$. Kubo paring determines static linear response: if the Hamiltonian is perturbed by $\lambda B$, where $\lambda$ is infinitesimal, then the change in the equilibrium expectation value of $A$ is
\begin{equation}
\Delta\langle A\rangle=\langle \Delta A\rangle-\beta\lambda\langle\langle  A;B\rangle\rangle +O(\lambda^2).
\end{equation}
Here the first term is due to the possible explicit dependence of $A$ on the Hamiltonian, while the second term is the change in the expectation value of $A$ due to the change in the equilibrium state.

Kubo pairing is symmetric, $\langle\langle A;B\rangle\rangle=\langle\langle B;A\rangle\rangle$, and satisfies
\begin{equation}
\beta\langle\langle i[H,A];B\rangle\rangle=\langle i[B,A]\rangle.
\end{equation}
In finite volume, one can write it in terms of the energy eigenstates as follows:
\begin{equation}
\langle\langle A;B\rangle\rangle=Z^{-1}\sum_{n,m} \langle n|\bar A|m\rangle\langle m|\bar B|n\rangle \frac{e^{-\beta E_m}-e^{-\beta E_n}}{\beta(E_n-E_m)},
\end{equation}
where $\bar A=A-\langle A\rangle,$ and $\bar B=B-\langle B\rangle.$

\section{Some mathematical constructions}\label{appendix:math}

\subsection{Chains and cochains}\label{appendix:chains}

In this paper we have encountered local operators $H_p$ and $Q_p$ which depend on a lattice point $p\in\Lambda$, operators $J_{pq}$ and $J^E_{pq}$ which depend on a pair of points, and energy magnetization $M^E_{pqr}$ which depends on three points. It is useful to introduce a suitable terminology for such objects. Let $n$ be a non-negative integer. Consider a quantity $A(p_0,\ldots,p_n)$ which depends on $n+1$ points of $\Lambda$, is skew-symmetric under the exchange of points, and decays rapidly  when the distance between any two points becomes large. Given an ordered set of points $p_0,\ldots,p_n\in\Lambda$, let $[p_0,\ldots,p_n]$ denote an abstract oriented $n$-simplex with vertices $p_0,\ldots,p_n$. Then one can consider a formal linear combination of simplices
\begin{equation}
A=\frac{1}{(n+1)!}\sum_{p_0,\ldots,p_n} A(p_0,\ldots ,p_n)[p_0,\ldots,p_n].
\end{equation}
Such a linear combination is called an $n$-chain, or a chain of degree $n$. For example, the operators $J_{pq}$ form an operator-valued 1-chain $J$. 

The simplest decay condition one can impose is to require $A(p_0,\ldots,p_n)$ to vanish whenever any two of its arguments are separated by more than some finite distance $\eps$. This distance may be different for different chains. In the body of the paper such chains are called finite-range. In the mathematical literature they are called controlled chains \cite{Roe}. The current 1-chain $J$ is finite-range, or controlled, because $[H_p,Q_q]=0$ for $|p-q|>\R.$ 

Another natural decay condition is to require $A(p_0,\ldots,p_n)$ to satisfy 
\begin{equation}
\sum_{p_0, p_1,\ldots,p_n}||A(p_0,\ldots,p_n)||<\infty .
\end{equation}
Here $||\cdot ||$ denotes operator norm if $A$ is operator-valued and absolute value if $A$ is real-valued or complex-valued.
We will call such chains summable. For example, the real-valued 2-chain defined in eq. (\ref{mue}) is summable if $\frac{dH_p}{d\lambda}$ is nonzero only for $p$ in a finite subset and the Kubo pairings of local operators decay rapidly with distance. 

There is an operation $\partial$ on chains which lowers the degree by $1$:
\begin{equation}
(\partial A)(p_1,\ldots,p_n)=\sum_{q\in\Lambda} A(q,p_1,\ldots,p_n).
\end{equation}
Although the sum is infinite, the operation  is well-defined for $n>0$ since we assumed rapid decay when $q$ is far away from any of the points $p_1,\ldots, p_n$. This operation  satisfies $\partial\circ\partial=0$. It maps controlled chains to controlled chains, and summable chains to summable chains. The chain $\partial A$ is called the boundary of the chain $A$. A cycle is a chain whose boundary is zero. Using this notation, the conservation equation (\ref{electricconserv}) can be written as 
\begin{equation}\label{conservation}
\frac{d Q}{dt}=-\partial J.
\end{equation}

Dually, an $n$-cochain is a function of $n+1$ points of $\Lambda$ which is skew-symmetric, but need not decay when one of the points is far from the rest. We will only consider real-valued cochains. A natural operation on cochains is:
\begin{equation}
(\delta \alpha)(p_0,\ldots,p_{n+1})=\sum_{j=0}^{n+1} (-1)^j \alpha(p_0,\ldots,p_{j-1},p_{j+1},\ldots,p_{n+1}).
\end{equation}
It increases the degree by $1$ and satisfies $\delta\circ\delta=0$. The cochain $\delta\alpha$ is called the coboundary of the cochain $\alpha$. A cocycle is a cochain whose coboundary is zero. The evaluation of an $n$-chain $A$ on an $n$-cochain $\alpha$ is formally defined as
\begin{equation}\label{value of chain on cochain}
A(\alpha)=\frac{1}{(n+1)!}\sum_{p_0,\ldots,p_n} A(p_0,\ldots,p_n)\alpha(p_0,\ldots,p_n).
\end{equation}
This definition is formal because without some constraints on the cochain $\alpha$ the infinite sum will not be absolutely converging.
An example of a 1-cochain is a function $\eta(p,q)$ which appears in (\ref{jf}), then the operator $J(\eta)$ is simply the evaluation of the operator-valued 1-chain $J$ on a 1-cochain $\eta$. 

Suppose all our chains are controlled. Then the problematic contribution in (\ref{value of chain on cochain}) arises from the region where all points $p_0,\ldots,p_n$ are nearby but otherwise can be anywhere in $\Lambda$. We will call such a region in the $(n+1)$-fold Cartesian product of $\Lambda$ with itself a thickened diagonal. To make the evaluation well-defined, it is natural to impose the following requirement on $\alpha$: the intersection of the support of $\alpha$ with any thickened diagonal must be finite. In the mathematical literature such cochains are called cocontrolled \cite{Roe}. For example, if we regard $\chi_B$ as a 0-cochain, then $\eta=\delta \chi_B$ is cocontrolled if either $A$ or $B$ are compact. One can evaluate an arbitrary complex-valued controlled $n$-chain on a cocontrolled $n$-cochain and get a well-defined number. Or, when one evaluates an operator-valued controlled chain on a cocontrolled cochain, one gets a bounded operator.

If our chains are summable, then it is natural to require $n$-cochains to be bounded functions on the $(n+1)$-fold Cartesian product of $\Lambda$ with itself.
The space of bounded cochains is the Banach-dual of the space of summable chains, where the norms are the obvious ones. Thus the evaluation of a chain on a cochain is well-defined and is a continuous function of both the chain and the cochain. 

With this said, we can state a kind of "Stokes' theorem"
\begin{equation}\label{stockes theorem}
A(\delta\beta)=\partial A(\beta).
\end{equation}
It applies to any controlled $n$-chain $A$ and any cocontrolled $(n-1)$-cochain $\beta$. It also applies to any summable $n$-chain and a bounded $(n-1)$-cochain. In the special case $A=J$ and $\beta=\chi_B$ for some finite set $B$, combining (\ref{stockes theorem}) and the conservation equation (\ref{conservation}) we get that the current through the boundary of $B$ (represented by the 1-cocycle $\delta \chi_B$) is equal to minus the rate of change of the total charge in $B$.

Given an $n$-cochain $\alpha$ and an $m$-cochain $\gamma$ one can define an $(n+m)$-cochain $\alpha\cup\gamma$ by 
\begin{align}
(\alpha\cup\gamma)(p_0,\ldots,p_{n+m})=\frac{1}{(n+m+1)!}\sum_{\sigma\in {\mathcal S}_{n+m+1}} (-1)^{{\rm sgn}\, \sigma} \alpha(p_{\sigma(0)},\ldots,p_{\sigma(n)})\gamma(p_{\sigma(n)},\ldots,p_{\sigma(n+m)}).
\end{align}
where ${\mathcal S}_{n+m+1}$ is the permutation group on $n+m+1$ objects. This operation  satisfies 
\begin{equation}
\alpha\cup\gamma=(-1)^{nm}\gamma\cup\alpha,\quad \delta(\alpha\cup\gamma)=\delta\alpha\cup\gamma+(-1)^n\alpha\cup\delta\gamma.
\end{equation}
The operations $\delta$ and $\cup$ on cochains are analogous to operations $d$ and $\wedge$ on differential forms. In the body of the paper we apply these formulas in the case when $\alpha=\delta f$ and $\gamma=\delta g$, where $f$ is a "smeared step-function" in the $x$-direction, and $g$ is a "smeared step-function" in the $y$-direction. The chains $\alpha$ and $\gamma$ are bounded, and so is $\alpha\cup\gamma$. Hence if $\frac{dH_p}{d\lambda}$ is nonzero only for a finite subset of $\Lambda$, the evaluation of the summable chain (\ref{mue}) on the bounded cochain $\delta f\cup\delta g$. More generally, we may consider uniform deformations such that $\frac{dH_p}{d\lambda}$ is bounded, but does not vanish at infinity. Then the chain (\ref{mue}) is only locally summable. Nevertheless, its evaluation on $\delta f\cup\delta g$ is still well-defined because $\delta f\cup\delta g$ is cocontrolled as well as bounded. 

Finally, we note that if an $n$-chain $A(p_0,\ldots,p_n)$ is nonzero only if $|p_i-p_j|\leq\delta$ for all $i,j$, then its contraction with an $n$-cochain $\alpha$ is well-defined even if $\alpha(p_0,\ldots,p_n)$ is only defined for $|p_i-p_j|\leq\delta$. 
We will make occasional use of such partially-defined cochains below.

\subsection{Applications}\label{appendix:applications}

In this section we discuss some physical application of the machinery of chains and cochains. As discussed above, electric current is a operator-valued controlled 1-chain satisfying (\ref{conservation}). A natural solution is given by (\ref{electric current}), but there is an obvious ambiguity (\ref{electric_current_ambiguity}). In the language of chains, it amounts to $J\mapsto J+\partial U$, where $U$ is an operator-valued controlled 2-chain. This ambiguity does not affect quantities like $J(\eta)$, where $\eta$ is a cocontrolled 1-cycle. Indeed, using the Stokes' theorem, we get $(J+\partial U)(\eta)=J(\eta)+U(\delta \eta)=J(\eta).$ Similarly, while the energy current (\ref{energycurrent}) has an obvious ambiguity (\ref{energy_current_ambiguity}), it does not affect quantities like $J^E(\eta)$, where $\eta$ is a cocontrolled 1-cocycle. A special case of this is the electric or energy current from region $B$ to region $A$ which is denoted $J(A,B)$ or  $J^E(A,B)$ in the body of the paper. This is a physically measurable quantity, and it is satisfying that is not affected by this ambiguity.

A more subtle question is whether there are other ambiguities in the definition of currents. This is equivalent to asking whether the equation $\partial \Delta J=0$ has solutions other than $\Delta J=\partial U$. To answer this question we need to know the homology of the complex of controlled chains in degree $1$. More generally, one might want to know the homology of the complex of controlled chains in all degrees. It turns out that under natural assumptions on the lattice $\Lambda\subset\RR^d$ the controlled homology in degree $n$ is independent of $\Lambda$ and equal to the locally-finite (Borel-Moore) homology of $\RR^d$ \cite{HigsonRoe}. The latter is equal to $0$ for $n\neq d$ and isomorphic to $\RR$ for $n=d$. The condition on $\Lambda$ is, roughly speaking, that it fills the whole $\RR^d$ uniformly. More precisely, there should exist a number $r>0$ such that any point of $\RR^d$ is within distance $r$ of some point of $\Lambda$. In the terminology of \cite{Roe}, this implies that $\Lambda$ is coarsely equivalent to $\RR^d$. 

Given this result, we see that for $d>1$ the only solutions to $\partial\Delta J=0$ have the form $\Delta J=\partial U$, where $U$ is a controlled operator-valued chain. In other words, our formulas for $J_{pq}$ and $J^E_{pq}$ are essentially unique. The case $d=1$ is a bit different, since the degree 1 homology of controlled chains is nontrivial. In the case $d=1$ points of $\Lambda$ can be naturally labeled by integers, and a nontrivial solution to $\partial\Delta J=0$ has the form $\Delta J_{pq}=J_0 (\delta(p,q-1)-\delta(q,p-1))$, where $J_0$ is a fixed local operator. However, if we make a natural assumption that $J_0$ must be supported in some fixed-size neighborhood of the points $p,q$ for all $p,q$, then $J_0$ must be proportional to the identity operator. The same applies to the energy current. Thus for $d=1$ system currents are unique up to an addition of a constant c-number. This c-number, if present, would violate the conclusion of Bloch's theorem \cite{Bloch} or its energy counterpart \cite{energyBloch}. It would lead to an unphysical electric or energy current even at $T=\infty$, when all degrees of freedom are in a maximally-mixed state. If we normalize the currents so that their  expectation values vanish at $T=\infty$, we eliminate this ambiguity even for $d=1$. With this normalization, both Bloch's theorem and its energy counterpart hold for all temperatures.

Another application is the definition of magnetization and energy magnetization. The equilibrium expectation value of the electric current satisfies 
\begin{equation}\label{conservation chain form}
\partial\langle J \rangle=0.
\end{equation}
An obvious solution has the form
\begin{equation}\label{magnetization}
\langle J\rangle=\partial M,
\end{equation}
where $M$ is a real-valued 2-chain. This is a lattice analog of of the continuum equation
\begin{equation}
\langle J_k(\br)\rangle=-\eps_{kj}\partial_j M(\br)
\end{equation}
which defines magnetization $M(\br)$. Thus one can regard the real-valued 2-chain $M_{pqr}$ as a lattice analog of magnetization.

In order for the magnetization 2-chain to exist, eq. (\ref{magnetization}) must be the most general solution of (\ref{conservation chain form}). Thus magnetization exists if the homology of $\partial$ in degree $1$ is trivial, or more generally, if the homology class of the 1-chain $\langle J\rangle$ is zero. Since the 1-chain $\langle J_{pq}\rangle$ is controlled, it is sufficient to look at the homology of controlled chains. If $\Lambda$ is coarsely equivalent to $\RR^d$ and $d>1$, the controlled homology in degree 1 is trivial, as explained above. Thus magnetization exists. It is not unique, of course, since there is always an ambiguity $M\mapsto M+\partial P,$ where $P$ is any real-valued controlled 3-chain. This is a harmless ambiguity since physical expressions involve expressions like $M(\zeta)$, where $\zeta$ is a cocontrolled 2-cochain and are unaffected. A more serious ambiguity arises if controlled homology of $\Lambda$ in degree 2 is non-trivial. This is the case if $\Lambda$ is coarsely equivalent to $\RR^2$. Given any magnetization 2-chain, one can get another acceptable magnetization 2-chain by adding to it a controlled 2-cycle. Thus magnetization has an unavoidable ambiguity for 2d lattices, but not for lattices of higher dimensions. The same remarks apply verbatim to energy magnetization.

The case $d=1$ is again a bit special. Controlled homology in degree 1 is nontrivial, but $\langle J(\eta)\rangle=0$ for any cocontrolled 1-cochain thanks to Bloch's theorem. Hence the homology class of $\langle J\rangle$ is trivial, and magnetization still exists. The same applies to the energy current and energy magnetization.

Finally, the homology of summable chains is trivial in degree higher than 0 for any lattice $\Lambda$. This is proved by exhibiting a contracting 
homotopy for the summable chain complex. Therefore if $\frac{dH_p}{d\lambda}$ is supported on a finite set, the chain (\ref{mue}) 
is unique up to a replacement $\mu^E\mapsto\mu^E+\partial N$, where $N$ is a summable 3-cochain. This shows that our expression for $\mu^E$ is essentially unique for deformations of the Hamiltonian which are  supported on a finite set. Since a general bounded deformation can be written as an (infinite) sum of these, we conclude that our formula for $\mu^E$ is essentially unique.

\section{Exponential decay of certain correlators in a gapped phase}

\label{appendix: watanabe}

Let $A$, $B$, and $C$ be local operators such that the supports of $A$ and $B$ are separated by at least $L$. Let $G=(z-H)^{-1}$ be the Green's function of a gapped Hamiltonian, and let $E_0$ be the energy of the ground state. For the time being we assume that the ground state is unique and comment on the more general case later. We are going to prove that the correlator
\begin{align}
\begin{split}
  \oint_{z=E_0} \frac {dz}{2\pi i} {\rm Tr \,} \left( \left[A, G  BG\right] G C G\right),
\end{split}
\end{align}
is exponentially suppressed for large $L$.  Note that the support of the operator $C$ is not required to be separated from the supports of $A$ and $B$.  By performing the $z$ integration we get
\begin{align}\label{expanded correlator}
\begin{split}
  \oint \frac {dz}{2\pi i} {\rm Tr \,} \left( \left[A, G  BG\right] G C G\right)= \langle A G_0 B G_0^2 C \rangle +\langle B G_0^2 C G_0 A \rangle - \langle C G_0^2 AG_0 B \rangle \\- \langle C G_0 A G_0^2 B\rangle+\langle B G_0^2A G_0 C \rangle + \langle B G_0 A G_0^2 C \rangle - \langle A G_0 C G_0^2 B \rangle\\ - \langle C G_0^2 B G_0 A \rangle   + 2 \left(\langle  C G_0^3 B\rangle- \langle B G_0^3 C \rangle \right) \langle A \rangle\\ + \Big( \langle A G_0^3B\rangle-\langle B G_0^3A\rangle\Big) \langle C \rangle+ \left(\langle C G_0^3 A \rangle - \langle A G_0^3 C  \rangle\right)\langle B\rangle,
\end{split}
\end{align}
where $\langle\ldots\rangle$ denotes the average over the ground state and we have introduced the notation
\begin{align}
G_0 = \sum_{n\ne0} \frac{|n\rangle\langle n|}{E_0-E_n}.
\end{align}

Now we use the following facts from \cite{Watanabe} and other similar identities:
\begin{align}\label{Watanabe formuals}
\begin{split}
\langle O_1 G_0^n O_2 G_0^m O_3 \rangle &= \langle O_1 G_0^{n+m}  O_3 \rangle \langle O_2\rangle+O(e^{-L/\xi}),\\
\langle O_2 G_0^n O_1 G_0^m O_3 \rangle &=O(e^{-L/\xi}),\\
\langle O_1 G_0^n O_2\rangle &=O(e^{-L/\xi}),
\end{split}
\end{align}
if $n,m>0$ and the support of operator $O_2$ is at least $L$ distance away from the supports of $O_1$ and $O_3$. Here $\xi>0$ is a scale parameter which is finite for gapped systems. See \cite{Watanabe} for the derivation of these identities. 

Using these we can simplify the first term in (\ref{expanded correlator}). Separating $C$ (which is by assumption a sum of local operators) into two parts $C= C_A+ C_B$ where the support of $C_A$ is far away from $B$ and the support of $C_B$ is far away from $A$, we get

\begin{align}
\begin{split}
 \langle A G_0 B G_0^2 C\rangle &= \langle A G_0 B G_0^2 C_A \rangle+\langle A G_0 B G_0^2 C_B  \rangle \\ &= \langle A G_0^3 C_A  \rangle\langle  B\rangle +O(e^{-L/\xi})=  \langle A G_0^3 C \rangle\langle  B\rangle+O(e^{-L/\xi}).
\end{split}
\end{align}

Similarly, we have
\begin{align}
 \langle B G_0^2 C G_0 A \rangle &= \langle B G_0^3 A \rangle \langle C \rangle +O(e^{-L/\xi}), \\
  - \langle C G_0^2 AG_0 B \rangle&=- \langle C G_0^3 B \rangle \langle A\rangle +O(e^{-L/\xi}),\\
  - \langle C G_0 A G_0^2 B\rangle&=  - \langle C G_0^3 B\rangle\langle A \rangle +O(e^{-L/\xi}),\\
  \langle B G_0^2A G_0 C \rangle&= \langle B G_0^3 C\rangle\langle A \rangle +O(e^{-L/\xi}),\\
  \langle B G_0 A G_0^2 C \rangle &= \langle B G_0^3 C \rangle\langle A \rangle +O(e^{-L/\xi}),\\
  -\langle A G_0 C G_0^2 B \rangle&= -\langle A G_0^3 B \rangle\langle C\rangle +O(e^{-L/\xi}),\\
  - \langle C G_0^2 B G_0 A \rangle&= - \langle C G_0^3 A \rangle\langle B \rangle +O(e^{-L/\xi}).
\end{align}
These eight terms exactly cancel the remaining six terms in (\ref{expanded correlator}). Putting everything together, we get
\begin{align}
\begin{split}
  \oint_{z=E_0} \frac {dz}{2\pi i} {\rm Tr \,} \left( \left[A, G  BG\right] G C G\right)=O(e^{-L/\xi}).
\end{split}
\end{align}

We have assumed a single ground state in the above derivation. However, as noted in \cite{Watanabe}, exactly the same arguments work for a $q$-fold degenerate ground state assuming that they are indistinguishable by local operators, i.e. if
\begin{equation}
\langle p | O|q \rangle  = \delta_{pq} \langle p | O|p\rangle + O(L^{-\infty}) 
\end{equation}
where $|p\rangle,|q\rangle$ are ground states, $O$ is a local operator, and $L$ is the size of the system.

\section{On the path-independence of the relative thermal Hall conductance}\label{appendix:path}

In this section we give a more detailed argument showing that the the relative thermal Hall conductance is independent of the choice of the path connecting two points in the parameter space of 2d systems with finite correlation length. As explained in the body of the paper, it is sufficient to show that the 1-form $\mu^E(\delta f\cup\delta g)$ is exact. Here $f(p)$ and $g(p)$ are smeared step-functions in the $x$ and $y$ directions. Let $f(p)$ be constant except for $x(p)\simeq a$, and $g(p)$ be constant except for $y(p)\simeq b$.

The first step is to make the $y$ direction periodic with period $L$, thereby replacing $\RR^2$ with a cylinder $\RR_x\times S^1_y$. For $L$ much larger than the correlation length this will change local quantities such as $\mu^E_{pqr}$ by an amount of order $L^{-\infty}$. 
One complication is that the function $g(p)$ is not periodic in the $y$ direction and thus does not descend to $\RR_x\times S^1_y$. We deal with this by reinterpreting $(\delta g)(p,q)=g(q)-g(p)$ as a function on $\Lambda\times\Lambda$ defined only for $|p-q|<L/2$. To make the evaluation of $\mu^E$ on $\delta f\cup\delta g$ well-defined, we truncate $\mu^E_{pqr}$ to zero whenever any two of the points $p,q,r$ are farther apart than $L/2$. Let us denote the truncated energy magnetization by $\tilde\mu^E_{pqr}.$ Because of truncation, we now have $d\langle J_{pq}\rangle=\sum_r \tilde\mu^E_{pqr}+O(L^{-\infty})$. Or using the notation of Appendix \ref{appendix:math},
\begin{equation}
d\langle J\rangle =\partial \tilde\mu^E+O(L^{-\infty}).
\end{equation}
Naively, one can deduce the desired result using the Stokes' theorem (\ref{stockes theorem}):
\begin{equation}
\int_\Gamma \tilde\mu^E(\delta  f\cup\delta g) =\int_\Gamma d\langle J^E\rangle (f\cup\delta g)+O(L^{-\infty})=O(L^{-\infty}).
\end{equation}
This argument is not correct because the 1-cochain $f\cup \delta g$ is not cocontrolled (because $(f\cup\delta g)(p,q)$ does not vanish when $x(p)\simeq x(q)$ and both $x(p)$ and $x(q)$ are large and negative), and the evaluation of $d\langle J^E\rangle$ on such a 1-cochain is not well-defined. To fix this, we first modify the Hamiltonian for $x<a-L$ by scaling it to zero. Since there are no phase transitions in 1d systems, the correlation length remains finite, and therefore the effect of such a modification on $\tilde\mu^E(\delta f\cup\delta g)$ will be of order $L^{-\infty}$. Then the operator-valued chain $J^E$ also becomes zero for $x\ll a$, and the application of the Stokes' theorem becomes legitimate. This concludes the argument.

Since by definition $\mu^E(\delta f\cup\delta g)$ is the differential of energy magnetization in the neighborhood of the point $(a,b)$, this  result means that energy magnetization exists as a globally-defined function on the parameter space. This function is defined up to an additive constant.

\section{The low-temperature behavior of the 1-form $\Psi$ in a gapped system} \label{psi appednix}

In this appendix we analyze the properties of the 1-form $\Psi(f,g)$ whose integral defines the relative invariant of gapped 2d systems. We will have to use estimates on the behavior of certain correlation functions at low but non-zero temperature. More precisely, we will assume that if the $T\ra 0$ limit of a correlator is well-defined, then at sufficiently low temperature deviations from the $T=0$ value are of order $O(e^{-T^*/T})$ for some $T^*>0$.   Physically, this is what one expects for a Hamiltonian with a gap for localized excitations. 

One could try to prove it by putting the  system on a torus of finite size $L$. Then for a correlation function $C(T)$ one can construct a finite-size  analog $C(T,L)$ such that $C(T)=\lim_{L\ra\infty} C(T,L)$. The correlation function $C(T,L)$ can be rewritten in terms of many-body Green's function $G=(z-H)^{-1}$. For example, one can write 
\begin{equation}\int_0^\beta \langle A(-i\tau) B\rangle_L d\tau=Z^{-1}\oint e^{-\beta z} \frac{dz}{2\pi i} \Tr (G A G B), 
\end{equation}
where $Z$ is the partition function, and the contour surrounds all the eigenvalues of $H$. Now if we deform the contour into a pair of contours, one surrounding $z=E_0$ and the other surrounding all other eigenvalues, we see that for low $T$ the contribution of the first contour is exponentially close to its $T\ra 0$ limit, while the contribution of the second one is exponentially small at low $T$. Thus $C(T,L)-C(0,L)$ is exponentially small at low $T$. If we assume that the order of limits $T\ra 0$ and $L\ra\infty$ can be interchanged, we can conclude that $C(T)$ is exponentially small at low $T$. These arguments are at best heuristic, since it is far from clear when interchanging the order of limits is legitimate.

For simplicity of presentation we will work on $\RR^2$ and simply assume that correlation functions in gapped phase at non-zero temperature are exponentially closed to their zero-temperature expectation value. Also, we will consider the system at a  fixed non-zero temperature $T$ and will vary only the Hamiltonian. As was explained in Section \ref{sec:thermalHalltopinvariance}, rescaling the temperature is equivalent to rescaling the Hamiltonian. Finally, let us fix some $L>0$ which is much larger than the correlation length and define the $L$-support of a 1-cochain $\alpha$ to be the set of points $p\in\Lambda$ such that $\alpha(p,q)\neq 0$ for at least for one $q$ such that $|p-q|<L.$

Consider the integral of $\Psi(f,g)$ along a path connecting two zero-temperature phases $\ma$ and $\mb$:
\begin{align} 
\label{integral psi app}
I(\ma,\mb)= \int_{\ma}^\mb \Psi(f,g).
\end{align}
We will argue that it converges, does not change under the shift of the end points $\ma,\mb$ as long as they do not cross zero-temperature phase transitions, and does not change under suitable  deformations of $f,g$.

Let us start with the last property. We consider adding to $f$ a function of $x(p)$ which has compact support (as a function of $x$) . We need to show that 
\begin{align}
    \int_{\ma}^\mb \Psi(f_0,g) = 0,
\end{align}
where $f_0$ is as in Fig \ref{g hat fig}. Since the path in the parameter space is away from phase transitions, the correlation length is finite everywhere along the path. Truncating $f_0$ to zero a distance $L$ away from the $L$-support of $\delta g$ will introduce error of order $L^{-\infty}.$ Denote the truncated cochain $\widetilde f_0$. It has compact support, therefore we can rewrite the magnetization term as
\begin{align}\label{emterm}
\begin{split}
\mu^E \left( { \delta\widetilde{f_0}} \cup  {\delta  g} \right)&=\partial \mu^E\left( \widetilde{ f_0} \cup  {\delta  g} \right) = d\langle J^E(\widetilde{f_0} \cup \delta  g)\rangle = -\frac 1 2  d\langle i[H(\widetilde{f_0}),H(g)]\rangle ,
\end{split}
\end{align}
where in the last step we have used the  definition of $J^E$ and cup product.  The Kubo term, on the other hand, can be rewritten as
\begin{align}\label{C5}
\begin{split}
\kappa^{\rm Kubo}_{xy}(\widetilde{f_0},g) =- \beta^2\lim_{s\ra 0+}   \int_0^\infty  dt \,e^{-st} \langle \langle  \frac{dH(\widetilde{f_0},t)}{dt}; J^E(\delta  g)\rangle\rangle \\= \beta^2   \langle \langle H(\widetilde{f_0}); J^E(\delta  g)\rangle\rangle+\beta^2\lim_{s\ra 0+}s \int_0^{\infty}dt e^{-st}\langle \langle H(\widetilde{f_0},t); J^E(\delta  g)\rangle\rangle.
\end{split}
\end{align}
The last term is in general non-zero since $\langle \langle H(\widetilde{f_0},t); J^E(\delta  g)\rangle\rangle$ does not have to converge to zero as $t\rightarrow \infty$. However, at zero temperature and for a gapped Hamiltonian one can explicitly check that this term is zero. Indeed, expanding the expression in the energy eigenbasis we get
\begin{multline*}
    \lim_{s\ra 0+}s \int_0^{\infty}dt e^{-st}\langle \langle H(\widetilde{f_0},t); J^E(\delta  g)\rangle\rangle \\= -i\lim_{s\ra 0+}s\sum_{n>0}\frac{\langle 0 | H(\widetilde{f_0})  |n\rangle\langle n |J^E(\delta  g)|0\rangle  -\langle 0 |J^E(\delta  g) |n\rangle\langle n | H(\widetilde{f_0})  |0\rangle }{(E_0-E_n)^2}=0.
\end{multline*}
 Therefore at small but non-zero temperature we expect the second term in (\ref{C5}) to be exponentially suppressed. The remaining term can be rewritten as
\begin{align}
\begin{split}
\beta^2  d \langle \langle H(\widetilde{f_0}); J^E(\delta g)\rangle\rangle=\beta^2 d\langle \langle H(\widetilde{f_0}); -i[H,H(g)]\rangle\rangle=- \beta d\langle i [H(\widetilde{f_0}),H(g)]\rangle.
\end{split}
\end{align}
This term cancels the energy magnetization contribution (\ref{emterm}). Therefore $\Psi(\widetilde{f_0},g)$ is a differential of a function which is exponentially small for $T\ra 0$. Hence the integral of $\Psi(\widetilde{f_0},g)$ along a path connecting two gapped zero-temperature systems is zero. Therefore the integral of $\Psi(f_0,g)$ along the same path is of order $L^{-\infty}.$ Since $L$ is arbitrary, we can take the limit $L\ra\infty$ and conclude that the integral of $\Psi(f_0,g)$ along this path is zero. Similarly, one can prove that $I(\ma,\mb)$ does not change if we add to $g$ a compactly supported function of $y$.

It is tempting to use the same argument with $f_0$ replaced with $f$ to show  that $I(\ma,\mb)$ is zero. But the argument cannot be carried through because it is impossible to truncate $f$ and make its support compact in such a way that the support of $\delta f \cup \delta g$ coincides with the support of  $\delta \widetilde f \cup \delta g$. There will necessarily be additional intersections.

In order to show that the integral (\ref{integral psi app}) defining $I(\ma,\mb)$ converges and is independent of the precise choice of endpoints,  consider a variation of the Hamiltonian supported in a quadrant of $\RR^2$. A general perturbation can be decomposed into a sum of four such perturbations. As discussed in the body of the paper, in order to show that $I(\ma,\mb)$ is independent of endpoints and converges it is sufficient to show that all components of the 1-form $\Psi(f,g)$ are exponentially small as $T\ra 0$. Following the same logic as before, we can shift $f,g$ in $\Psi(f,g)$ away from the support of the variation introducing an error which is exponentially small in temperature.  Recall that the 1-form $\Psi$ is defined as
\begin{equation}\label{g(T)}
\Psi(f,g)=  \beta^2 \left[d\int_0^\infty \beta e^{-st}\langle\langle J^E(\da,t);J^E(\dg)\rangle\rangle dt  -2\mu^E(\da\cup\dg)\right].
\end{equation}
Using the same arguments as in Section \ref{sec:invariance of sigma}, one can show that expression in square brackets is zero at $T=0$. Therefore, it is  exponentially small at zero temperature, and the same applies to $\Psi(f,g)$.

\section{Free fermion systems}
\label{appednix:free fermions}
Consider a free fermionic system on a lattice with a Hamiltonian
\begin{equation}
H=\sum_{p,q} a^\dagger(p) h(p,q) a(q).
\end{equation}
The infinite matrix $h(p,q)$ is assumed Hermitian, $h(p,q)^*=h(q,p)$. The energy on site $p$ is taken to be
\begin{equation}
H_p=\frac12\sum_m \left(a^\dagger(p)h(p,m)a(m)+a^\dagger(m) h(m,p) a(p)\right).
\end{equation}
Defining the charge operator as a 0-chain 
\begin{equation}
    Q_p = a^\dagger(p)a(p),
\end{equation}
we find the electric current 1-chain
\begin{equation}
    J_{pq} = i (a^\dagger(q) h(q,p) a(p) - a^\dagger(p) h(p,q) a(q)).
\end{equation}
Contracting it with a 1-cochain $\alpha(q)-\alpha(p)$ for some function $\alpha:\Lambda\ra \RR$, we get 
\begin{equation}
J(\delta \alpha)=-i a^\dagger [h,\alpha] a,
\end{equation}
where we now regard $\alpha$ as an operator in the one-particle Hilbert space.

Similarly, the energy current operator is a 1-chain 
\begin{multline}
J^E_{pq}=\frac{-i}{4}\sum_m\left(a^\dagger(p)h(p,q)h(q,m) a(m)-a^\dagger(q) h(q,p)h(p,m) a(m)\right.\\
\left. -a^\dagger(m) h(m,q)h(q,p) a(q)+a^\dagger(m)h(m,p)h(p,q) a(q)\right.\\
\left. +a^\dagger(p)h(p,m)h(m,q)a(q)-a^\dagger(q)h(q,m)h(m,p)a(p)\right).
\end{multline}
Contracting it with a 1-cochain $\alpha(q)-\alpha(p)$, we get 
\begin{equation}
J^E(\delta \alpha)=-\frac{i}{2} a^\dagger [h^2,\alpha] a.
\end{equation}
The Gibbs state at temperature $T=1/\beta$ is defined via
\begin{eqnarray}
\langle a(p,t) a^\dagger (q,0)\rangle &= &\left\langle p\left\vert \frac{e^{-ih t}}{1+e^{-\beta h}}\right\vert q\right\rangle, \\
\langle a(p,t)^\dagger a (q,0)\rangle &= & \left\langle q\left\vert \frac{e^{ih t}}{1+e^{\beta h}}\right\vert p\right\rangle,
\end{eqnarray}
and Wick's theorem. Then
\begin{equation}
\langle J(\delta f,t) J(\delta g)\rangle=- {\rm Tr}\left( [h, f] \frac{e^{-iht}}{1+e^{-\beta h}} [h, g] \frac{e^{iht}}{1+e^{\beta h}}\right),
\end{equation}
where the trace on the r.h.s. is taken over the 1-particle Hilbert space $L^2(\Lambda)$, and the functions $ f:\Lambda\ra \RR$ and $ g:\Lambda\ra \RR$ are regarded as Hermitian operators on this Hilbert space. The operators $[h, f]$ and $[h, g]$ are supported on a vertical and  a horizontal strips, respectively. 

Going to the energy basis, replacing $t\ra t-i\tau$ and integrating over $\tau$ from $0$ to $\beta$ we get
\begin{equation}
\langle\langle J(\delta f,t); J(\delta g)\rangle\rangle=\frac{-1}{\beta} \sum_{n,m} \langle n\vert [h, f]\vert m\rangle \langle m\vert [h, g]\vert n\rangle e^{i(\varepsilon_n-\varepsilon_m) t} \frac{e^{\beta \varepsilon_n}-e^{\beta \varepsilon_m}}{(1+e^{\beta \varepsilon_n})(1+e^{\beta \varepsilon_m})(\varepsilon_n-\varepsilon_m)},
\end{equation}
where $\varepsilon_n$ are 1-particle energy levels.
Note that in the limit $T\ra 0$, the fraction in this equation is equal to $\frac{\theta(\varepsilon_n)-\theta(\varepsilon_m)}{\varepsilon_n-\varepsilon_m}$ plus exponentially small terms. Thus at zero temperature $\varepsilon_m$ and $\varepsilon_n$ must have opposite signs. More generally, we can re-write the fraction as
\begin{equation}
\frac{\mathfrak f(\varepsilon_m)-\mathfrak f(\varepsilon_n)}{\varepsilon_n-\varepsilon_m}
\end{equation}
where $\mathfrak  f(\varepsilon)=\frac{1}{1+e^{\beta \varepsilon}}$ is the Fermi-Dirac distribution.

Integrating over $t$, we get
\begin{equation}
\sigma( f, g)=i \sum_{n,m} \frac{\langle n\vert [h, f]\vert m\rangle \langle m\vert [h, g]\vert n\rangle}{\varepsilon_n-\varepsilon_m+is} \frac{\mathfrak f(\varepsilon_n)-\mathfrak f(\varepsilon_m)}{\varepsilon_n-\varepsilon_m} 
\end{equation}

It is convenient to rewrite this expression using the 1-particle Green's functions $G_\pm(z)=1/(z-h\pm i0)$. The following formulas are useful:
\begin{equation}
\langle a^\dagger A a\rangle=-\frac 1 {2\pi i}\int_{-\infty}^\infty dz \, \mathfrak f(z) \Tr  \Big(\big[G_+-G_-\big] A\Big),
\end{equation}
\begin{align}
    \begin{split}
      -\beta\langle\langle a^\dagger A a; a^\dagger B a\rangle\rangle=-\frac 1 {2\pi i}\int_{-\infty}^\infty dz \mathfrak f(z)\Tr \Big(\big[G_+-G_-\big] A G_+ B + G_-  A\big[G_+-G_-\big] B\Big) \\=  -\frac 1 {2\pi i}\int_{-\infty}^\infty dz \mathfrak f(z)\Tr \Big(G_+ A G_+ B - G_-  A G_- B\Big) ,
    \end{split}
\end{align}
where we have suppressed the argument $z$ for $G_\pm(z)$. Here $A$ and $B$ are operators on the 1-particle Hilbert space and we have assumed $\langle a^\dagger A a\rangle=\langle a^\dagger B a\rangle =0$ in the last formula. Also note that
\begin{equation}
hG_{\pm}=G_{\pm}h=zG_{\pm}-1,\quad [G_\pm,A]=G_\pm[h,A]G_\pm.
\end{equation}
Using the Green's functions, the electric conductance can be rewritten as
\begin{multline}\label{free case sigma}
\sigma( f, g)=-\frac 1 {2\pi} \int_{-\infty}^{\infty} dz \,\mathfrak f(z) \Tr\big\{ [h, f] G_+^2 [h, g] (G_+-G_-)-[h, f](G_+-G_-)[h, g]  G_-^2 \big\},
\end{multline}
and the Kubo part of the thermal conductance as
\begin{multline}
\kappa^{\rm Kubo}_{xy}( f, g)=-\frac \beta {8\pi} \int_{-\infty}^{\infty} dz \, \mathfrak f(z) \Tr\big\{ [h^2, f] G_+^2 [h^2, g] (G_+-G_-)-[h^2, f](G_+-G_-)[h^2, g]  G_-^2 \big\}.
\end{multline}

The value of energy magnetization $\mu^E$ on a 2-cochain $\delta  f \cup \delta  g$ can be found to be 
\begin{multline}\label{variation of magnetization free system}
\mu^E(\delta f \cup \delta  g) = \frac 1 {16\pi}\int_{-\infty}^\infty dz \, \mathfrak f(z)\Tr\Big(  G_+dhG_+\Big\{  \big[[h, f],[h, g]\big]+[h^2, f]G_+[h, g]+[h, f]G_+[h^2, g]\\-[h^2, g]G_+[h, f]-[h, g]G_+[h^2, f]\Big\}\Big)
-(G_+ \rightarrow G_-),
\end{multline}
where $dh$ is the variation of the 1-particle Hamiltonian. In the translationally invariant case, one can replace $ f$ and $ g$ with momentum derivatives.

Using the above formulas, it is straightforward to compute the 1-form $\Psi$ for any free system.  Let us demonstrate this by computing the $T$-component of the 1-form $\Psi$.

For a global re-scaling of the Hamiltonian we have $dh=h$, and  eq. (\ref{variation of magnetization free system}) can be simplified
\begin{multline}
\tau^E(\delta f \cup \delta  g)=-\frac 1 {16\pi}\int_{-\infty}^\infty dz \, \Tr\Big\{2\mathfrak f(z) G_-^2 [h^2, f] (G_+-G_-) [h^2, g]\\-2\mathfrak f(z)  (G_+-G_-)  [h^2, f]G_+^2[h^2, g] +4\mathfrak f'(z) h^2 (G_+-G_-)  [h, f]G_+[h, g]\\-4\mathfrak f'(z)    G_-[h, f]h^2(G_+-G_-)[h, g]-\mathfrak f'(z) h (G_+-G_-)[[h, f],[h,\beta]] \Big\}.
\end{multline}

Variation of $\kappa^{\rm Kubo}_{xy}( f, g)$ contains two pieces:
\begin{multline}
   -\frac \beta {8\pi}  d\left(\int_{-\infty}^{\infty} dz \,\mathfrak f(z) \Tr\Big\{ [h^2, f] G_+^2 [h^2, g] (G_+-G_-)\Big\} \right)\\=  \frac \beta {8\pi} \int_{-\infty}^{\infty} dz \Tr\Big\{-2  \mathfrak f(z)[h^2, f] G_+^2 [h^2, g] (G_+-G_-)- 4 \mathfrak f'(z)[h, f] G_+^2 [h, g]h^3 (G_+-G_-)\\+4 \mathfrak f'(z)[h, f] G_+ [h, g]h^2 (G_+-G_-)- \mathfrak f'(z)[h, f] [h, g]h (G_+-G_-)\Big\}
\end{multline}
and
\begin{multline}
   \frac \beta {8\pi}  d\left(\int_{-\infty}^{\infty} dz\, \mathfrak f(z) \Tr\Big\{ [h^2, g] G_-^2 [h^2, f] (G_+-G_-)\Big\} \right)\\ =  \frac \beta {8\pi} \int_{-\infty}^{\infty} dz \Tr\Big\{2  \mathfrak f(z)[h^2, g] G_-^2 [h^2, f] (G_+-G_-)+ 4 \mathfrak f'(z)[h, g] G_-^2 [h, f]h^3 (G_+-G_-)\\-4 \mathfrak f'(z)[h, g] G_- [h, f]h^2 (G_+-G_-)+ \mathfrak f'(z)[h, g] [h, f]h (G_+-G_-)\Big\}.
\end{multline}

Inserting these three contributions into eq. (\ref{Psi n}) we arrive at
\begin{multline}
    \frac{d}{dT}\left(\frac{\kappa_{xy}( f, g)}{T}\right) =  \frac 1 {2\pi T^3} \int_{-\infty}^{\infty} dz \Tr\Big\{ \mathfrak f'(z)[h, f] G_+^2 [h, g]z^3 (G_+-G_-)\\- \mathfrak f'(z)[h, g] G_-^2 [h, f]z^3 (G_+-G_-)\Big\}.
\end{multline}
The right-hand side looks very similar to the electric conductance (\ref{free case sigma}). Indeed, integrating it over temperature from $0$ to $\infty$ and using the formula
\begin{align}
 \int_0^\infty \frac {dT} {T^3} \mathfrak f'(z) = -\frac  {\pi^2} {6 |z|^3} = -\frac{\pi^2}{3z^3}\left(\mathfrak f(z)\Big \vert_{T=\infty}-\mathfrak f(z)\Big \vert_{T=0}\right)
\end{align}
gives 
\begin{align}
    \frac{\kappa^A}{T} \Big \vert_{T=\infty} -  \frac{\kappa^A}{T} \Big \vert_{T=0} =\frac {\pi^2}3\left( \sigma^A\Big \vert_{T=\infty} -\sigma^A\Big \vert_{T=0}\right).
\end{align}
Since at infinite temperature the electric Hall conductance vanishes, while the thermal Hall conductance can be defined to vanish, we arrive at the Wiedemann-Franz law.

\end{document}